\documentclass[12pt]{article}
\usepackage{amsmath}
\usepackage{graphicx}
\usepackage{enumerate}
\usepackage{natbib}
\usepackage{url} 
\usepackage[colorlinks=true,allcolors=blue]{hyperref}
\usepackage{url}
\usepackage{graphicx}
\usepackage{amsmath, amssymb}
\usepackage{bbold}
\usepackage{mathbbol}
\usepackage{algorithm}
\usepackage{algorithmic}
\usepackage{enumitem}
\usepackage{float}
\usepackage{xcolor}
\usepackage{bm} 
\usepackage{setspace}
\usepackage{multirow}
\usepackage{lscape}
\usepackage{rotating}
\usepackage{adjustbox}
\usepackage{booktabs}
\usepackage{ragged2e}
\newtheorem{theorem}{Theorem}
\usepackage{caption}
\usepackage{subcaption}
\usepackage{float}
\def\*#1{\mathbf{#1}}
\def\+#1{\amsmathbb{#1}}
\def\^#1{\mathbb{#1}}
\DeclareSymbolFontAlphabet{\amsmathbb}{AMSb}%

\newcommand{\blind}{1}

\begin{document}

\def\spacingset#1{\renewcommand{\baselinestretch}%
{#1}\small\normalsize} \spacingset{1}


\if1\blind
{
  \title{\bf Distributional outcome regression via quantile functions and its application to modelling continuously monitored heart rate and physical activity}
  \author{Rahul Ghosal$^1$, Sujit K. Ghosh$^{2}$, Jennifer A. Schrack$^{3}$, Vadim Zipunnikov$^{4}$ \hspace{.2cm}\\
     $^{1}$ Department of Epidemiology and Biostatistics, University of South Carolina \\
$^{2}$Department of Statistics, North Carolina State University\\
$^{3}$ Department of Epidemiology,  Johns Hopkins Bloomberg \\School of Public Health\\
$^{4}$ Department of Biostatistics, Johns Hopkins Bloomberg \\School of Public Health \\}
  \maketitle
} \fi

\if0\blind
{
  \bigskip
  \bigskip
  \bigskip
  \begin{center}
    {\LARGE\bf Distributional outcome regression via quantile functions and its application to modelling continuously monitored heart rate and physical activity}
\end{center}
  \medskip
} \fi

\bigskip
\begin{abstract}
Modern clinical and epidemiological studies widely employ wearables to record parallel streams of real-time data on human physiology and behavior. With recent advances in distributional data analysis, these high-frequency data are now often treated as distributional observations resulting in novel regression settings. Motivated by these modelling setups, we develop a distributional outcome regression via quantile functions (DORQF) that expands existing literature with three key contributions: i) handling both scalar and distributional predictors, ii) ensuring jointly monotone regression structure without enforcing monotonicity on individual functional regression coefficients, iii) providing statistical inference via asymptotic projection-based joint confidence bands and a statistical test of global significance to quantify uncertainty of the estimated functional regression coefficients. The method is motivated by and applied to Actiheart component of Baltimore Longitudinal Study of Aging that collected one week of minute-level heart rate (HR) and physical activity (PA) data on 781 older adults to gain deeper understanding of age-related changes in daily life heart rate reserve, defined as a distribution of daily HR, while accounting for daily distribution of physical activity, age, gender, and body composition. Intriguingly, the results provide novel insights in epidemiology of daily life heart rate reserve.

\end{abstract}

\noindent%
\vfill

\newpage
\spacingset{1.9} 
\section{Introduction}
\label{sec:intro4}
 With the advent of digital health technologies and wearables, many studies collect parallel streams of high frequency data on human physiology and behaviour including heart rate, physical activity (such as steps, activity counts), continuously monitored blood glucose, and others. This paper is motivated by wearable data from Baltimore Longitudinal Study of Aging (BLSA), a study of normative human aging, established in 1958 and conducted by the National Institute of Aging Intramural Research Program. Actiheart component of BLSA collected one week of minute-level heart rate (HR) and physical activity (PA) data on 781 older adults using an Actiheart, a chest-worn heart rate and activity monitor \citep{schrack2018using}. 
 
\begin{figure}[H]
	\centering
 
\begin{subfigure}[b]{0.45\textwidth}
\centering\includegraphics[width=1\textwidth,height=0.99\textwidth]{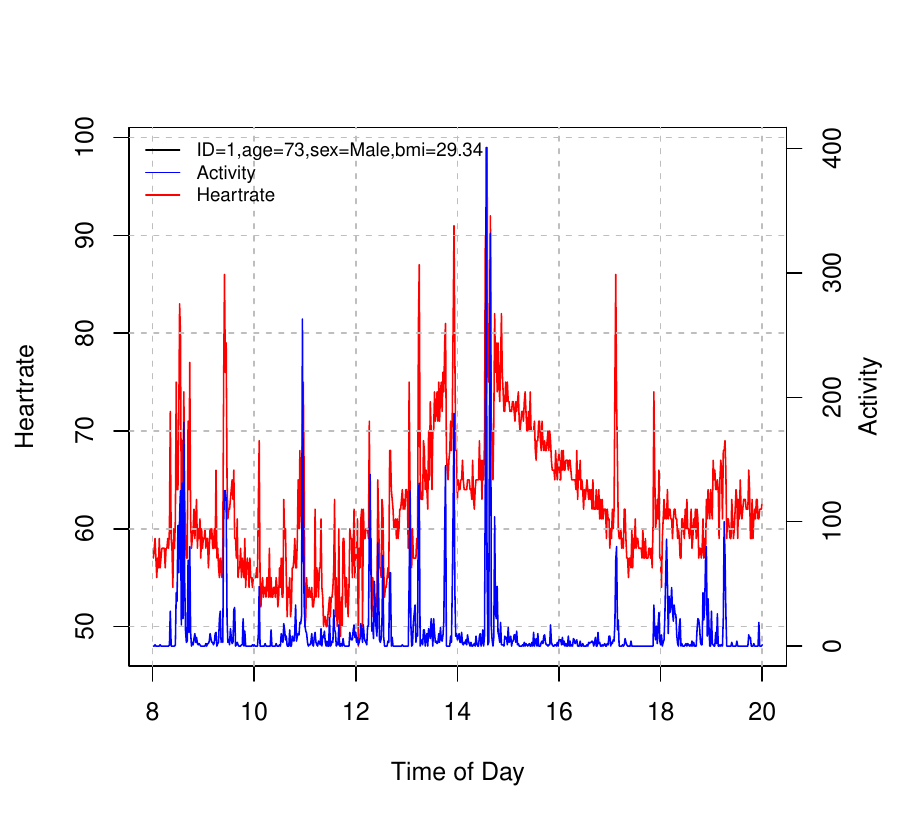}
\caption{Diurnal profile of heart rate and physical activity between 8 a.m.- 8 p.m. for a male subject (ID=1) in BLSA along with his age, gender and BMI.}
         \label{fig:t1}
     \end{subfigure}
    \hfill
 \begin{subfigure}[b]{0.45\textwidth}
         \centering
\includegraphics[width=1\textwidth,height=0.99\textwidth]{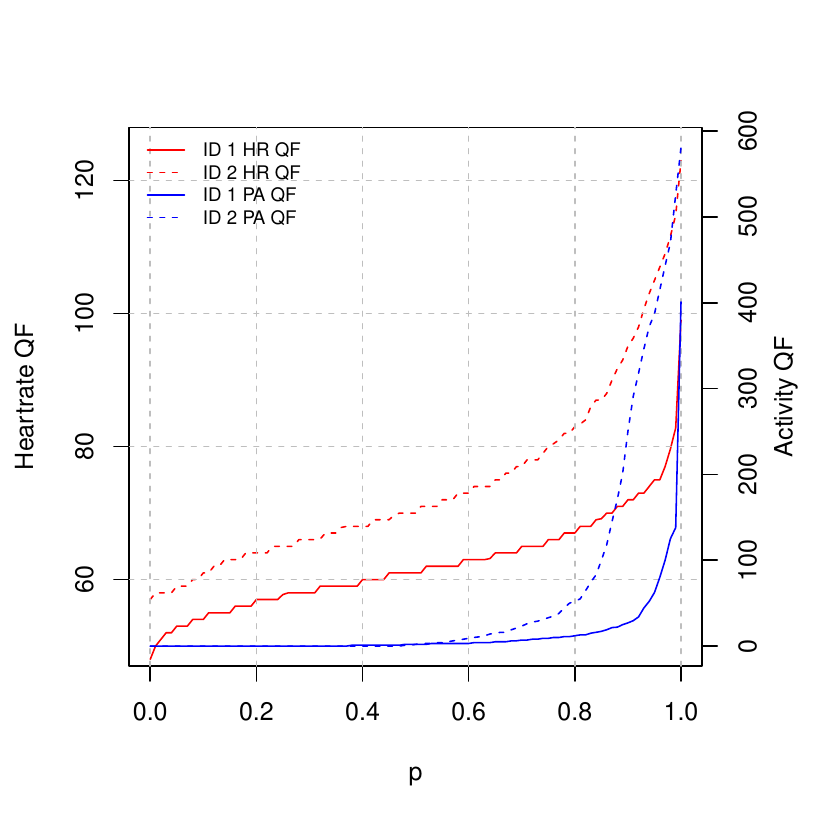}
\caption{Subject specific quantile functions of heart rate (HR) and physical activity (PA) for  two BLSA participants shown in (a) and (c).} 
\label{fig:t2}	
\end{subfigure}
\vfill
 \begin{subfigure}[b]{0.45\textwidth}
         \centering
\includegraphics[width=1\textwidth,height=0.99\textwidth]{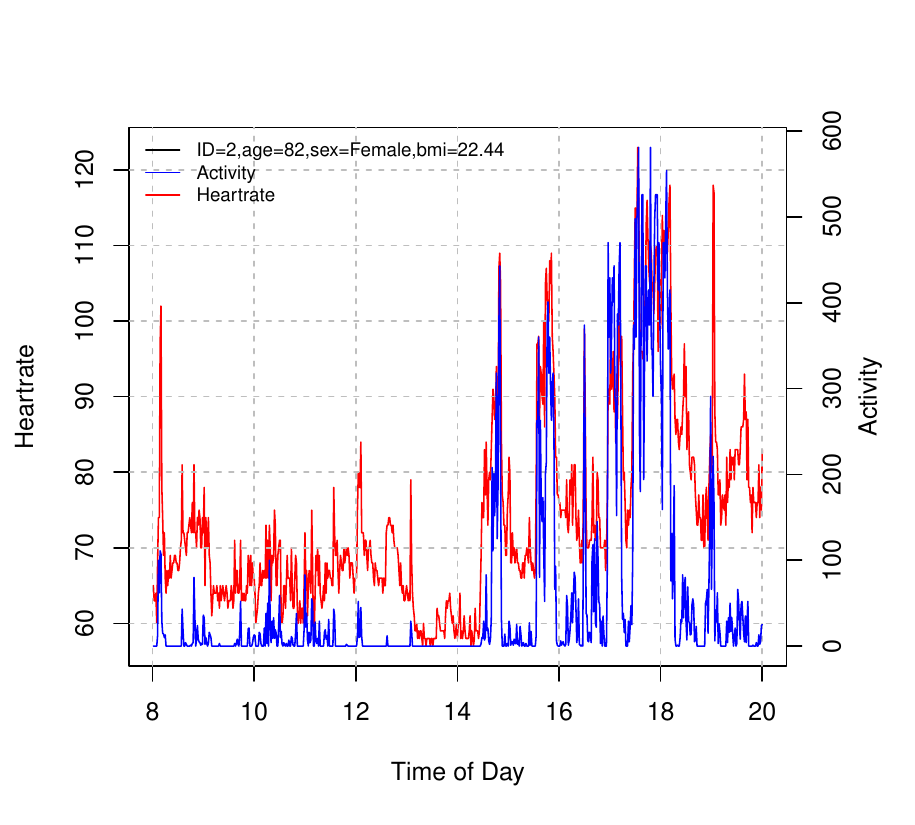}
\caption{Diurnal profile of heart rate and physical activity between 8 a.m.- 8 p.m. for a female subject (ID=2) in BLSA along with her age, gender and BMI.} 
\label{fig:t3}	
     \end{subfigure}    
 \hfill
\begin{subfigure}[b]{0.46\textwidth}
         \centering
\includegraphics[width=0.9\textwidth,height=0.85\textwidth]{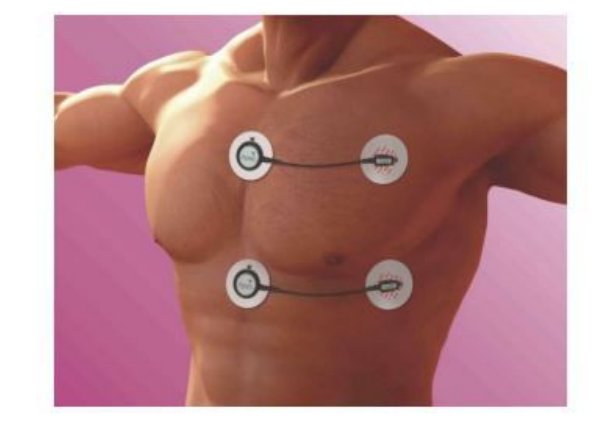}
\caption{Actiheart device (as in \citep{rautaharju1998standardized,ach}), used in BLSA for collection of heart rate (HR) and physical activity (PA).} 
\label{fig:t4}	
     \end{subfigure}
\caption{Illustration of Actiheart BLSA data collection with diurnal profiles and quantile functions of heart rate and physical activity.}
\label{Fig1}
\end{figure}

 Figure \ref{fig:t4} shows an Actiheart device and its typical placement on a chest (as in \cite{rautaharju1998standardized,ach}). Figures \ref{fig:t1} and \ref{fig:t3} display minute-level heart rate and minute-level activity counts (AC), unitless measures of physical activity intensity \citep{karas2022comparison}, during daytime (defined as 8AM-8PM) for a male and female BLSA participants. One of the main objective of BLSA study is to gain deeper understanding of age-related changes in aerobic and functional capacity of aging adults and effects of those changes on health and aging trajectory. One way to approach this question is to study age-related changes in daily life heart rate reserve (DL-HRR), that can be defined as a distribution of minute-level HR during typical wake time \citep{schrack2018using}. While modelling age-related changes in DL-HRR, it is important to account for factors that affect DL-HRR such as gender and body composition, quantified via body mass index (BMI), and for daily life composition of physical activity, quantified via the distribution of minute-level physical activity.  Figure \ref{fig:t2} shows DL-HRR with corresponding DL distributions of PA for two BLSA participants. 

 In this article, the primary motivation is to understand the age-related changes in ambulatory heart-rate (HR) during typical wake time, while accounting for gender, BMI and the daily physical activity (PA). There have been primarily two approaches for analysis of wearable data. The first approach comprises using various summary metrics of the data \citep{varma2018total,gait2020vr}.  Typically, specific scalar summaries of daily HR distribution such as resting HR, $Q_i(0)$ ($Q_i(\cdot)$ denoting quantile function), peak HR, $Q_i(1)$, or mean HR or median HR, $Q_i(0.5)$, are chosen and modelled as outcomes. While simplifying interpretability, this approach can suffer from loss of information when only considering the first few moments or specific summaries. If the research objective is to understand age-related changes in temporal/diurnal/circadian patterns of HR, functional data analysis methods \citep{goldsmith2016new,cui2022fast} would provide appropriate tools. Similarly, function-on-function and historical functional models could be used for modelling temporal heart rate profile as an outcome and temporal physical activity profile as the predictor. Distributional data analysis methods (DDA), on the other hand, provide modelling frameworks for capturing and modelling the distributional aspect of wearable data \citep{doi:10.1177/0962280221998064,ghosal2023distributional,matabuena2023distributional} using subject-specific distributional representations. Our modelling approach preserves entire daily distribution of HR by modelling subject-specific quantile function $Q_i(p), p\in [0,1]$ as the outcome (for example Figure 1 (b)). Thus, our approach preserves a substantial amount of original information in wearable data relevant to the scientific question we are motivated by. Note that, the goal here is to not model the raw data itself (which can be of huge volume), or it's temporal evolution, rather understand how different intensities of HR (e.g., maximal HR, HR reserve) depend on age, gender, BMI and subject-specific distribution of PA. In the BLSA application, we use multiple days of wearable data available for each study participant during their first visit, to estimate subject-specific quantile functions of heart rate and physical activity that characterize their entire distribution. The proposed modelling framework provided a novel way of understanding how the Wasserstein barycenter of the HR distribution change with changes in age, gender, BMI and distribution of PA. The illustrated data structure results in regression settings with distributional outcome (DL-HRR) and scalar (age, BMI, gender) and distributional predictors (DL distribution of PA). 

Below, we provide a brief review of the recent developments in distributional data analysis (DDA) and then discuss our proposal and contributions. The central idea of distributional data analysis is to capture the distributional aspects of subject-specific data and via histograms, densities, quantile functions and other distributional representations and use these as observations within various statistical modelling techniques.  \cite{petersen2021modeling} provided an in-depth overview of recent developments in DDA with a specific emphasis on using densities. Distributional data analysis has diverse applications across many scientific domain including digital health \citep{augustin2017modelling,matabuena2020glucodensities,ghosal2023distributional, matabuena2021distributional}, radiomics \citep{yang2020quantile}, neuroimaging \citep{tang2020differences} and many others. 

Similar to functional regression models, depending on whether the outcome or the predictor is distributional, there are various types of distributional regression models. \cite{petersen2016functional} and \cite{hron2016simplicial} developed functional compositional methods to analyze samples of densities. For scalar outcome and distributional predictors
the existing modelling approaches include scalar-on-quantile function regression \citep{ghosal2023distributional}, kernel-based approaches using quantile functions \citep{matabuena2023distributional}, density-transformation based approaches \citep{petersen2016functional,talska2021compositional} and many others (see \cite{petersen2021modeling}, \cite{chen2021wasserstein} and references therein).


In parallel, there has been a substantial work on developing models with distributional outcome and scalar predictors. \cite{yang2020quantile} developed a quantile function-on-scalar (QFOSR) regression model, where subject-specific quantile functions of data were modelled via scalar predictors of interest using a function-on-scalar regression approach \citep{Ramsay05functionaldata}, which make use of data-driven basis functions called quantlets. One limitation of the approach is a no guarantee of underlying monotonicity of the predicted quantile functions. To address this, \cite{yang2020random} extended this approach using I-splines \citep{ramsay1988monotone} or Beta CDFs which enforce monotonicity at the estimation step. One important limitation of this approach is enforcement of jointly monotone (non-decreasing) regression structure via enforcement of monotonicity on each individual functional regression coefficients. As it will be shown in BLSA Actiheart data, this assumption is too restrictive for two of our scalar predictors (age and gender).



Distribution-on-distribution regression models, when both outcome and predictors are distributions have been studied by \cite{irpino2013metric,chen2021wasserstein,ghodrati2022distribution,pegoraro2022projected}. These models aim to understand the association between distributions within a pre-specified, often linear, regression structure. \cite{irpino2013metric} used an ordinary least square approach modelling the outcome quantile function $Q_{iY}(p)$ as a non-negative linear combination of other quantile functions $Q_{iX_j}(p)$s. This approach is restrictive since it assumes a linear association between the distribution valued response and predictors, which are additionally assumed to be constant across all quantile levels $p \in (0, 1)$. \cite{chen2021wasserstein} used a geometric approach taking distributional valued outcome and predictor to a tangent space, where regular tools of function-on-function regression \citep{Ramsay05functionaldata,yao2005functional} were applied.
\cite{pegoraro2022projected} used an approximation of the Wasserstein space using monotone B-spline and developed methods for PCA and regression for distributional data. Recently, \cite{ghodrati2022distribution} developed a shape-constrained approach linking Frechet mean of the outcome distribution to the predictor distribution via an optimal transport map that was estimated by means of isotonic regression.

Many of above-mentioned methods mainly focused on dealing with constraints enforced by a specific functional representation. Developing inferential tools is somewhat under-developed are of distributional data analysis. \cite{chen2021wasserstein} derived the asymptotic convergence rates for the estimated regression operator in their proposed method for Wasserstein regression. \cite{yang2020quantile} developed joint credible bands for distributional effects, but monotonocity of the quantile function was not imposed. \cite{yang2020random} developed a global statistical test for estimated functional coefficients in the distributional outcome regression, however, no confidence bands was proposed to identify and test local quantile effects. 

In this paper, we propose a distributional outcome regression via quantile functions (DORQF) that provides three major contributions. First, DORQF includes both scalar and distributional predictors. Second, it ensures jointly monotone (non-decreasing) additive regression structure over the entire domain without enforcing monotonicity of individual functional regression coefficients. Third, it provides statistical inference tools for estimated functional regression coefficients including asymptotic projection-based joint confidence bands and a statistical test of global significance. We capture distributional aspect in outcome and predictors via quantile functions and construct a jointly monotone regression model via a specific shape-restricted functional regression model. The effect of scalar predictors is captured via functional coefficient $\beta_j(p)$'s varying over quantile levels and the effect of the distributional predictor is captured via a monotone function $h(\cdot)$, similar to an optimal transport approach in \cite{ghodrati2022distribution}. In the special case, when there is no distributional predictor, the model resembles a quantile function-on-scalar regression model, but with much less restrictive constraints than in \cite{yang2020random}. 
Bernstein polynomial (BP) basis functions are used to model the distributional effects $\beta_j(p)$s and the monotone map $h(\cdot)$, which are known to enjoy attractive and optimal shape-preserving properties \citep{lorentz2013bernstein,carnicer1993shape}. Additionally, BP is instrumental in constructing and enforcing a jointly monotone regression structure without over-restricting individual functional regression coefficients to be monotone. 



The rest of this article is organized as follows. We present our distributional modeling framework and illustrate the proposed estimation method in Section 2. In Section 3, we perform numerical simulations to evaluate the performance of the proposed method and provide comparisons with existing methods for distributional regression. In Section 4, we demonstrate application of the proposed method in modelling continuously monitored heart rate reserve in BLSA study. Section 5 concludes with a brief discussion of our proposed method and some possible extensions of this work.



\section{Methodology}
\label{sec:method1}
\subsection{Modelling Framework and Distributional Representations}
We consider the scenario, where there are repeated subject-specific measurements of a 
distributional response $Y$ (e.g., heart-rate in our motivating application) along with several scalar covariates $z_j$, $j=1,2,\ldots,q$ (e.g., age, gender, BMI) and we also have a distributional predictor $X$ (e.g., physical activity). Let us denote the subject-specific response and covariates as
$Y_{ik},X_{il},z_{ij} \hspace{1 mm}(k=1,\ldots,n_{1i}, l=1,\ldots,n_{2i})$, for subject $i=1,\ldots,n$. Here $n_{1i},n_{2i}$ denotes the number of repeated observations of the distributional response and predictor respectively for subject $i$.
The observed data can be represented as $D_i=\{Y_{ik}, X_{il},\*z_i;k=1,\ldots,n_{1i};l=1,\ldots,n_{2i}\}$, for subject $i=1,\ldots,n$, where $\*z_i=(z_{i1},\ldots z_{iq})$. Assume $Y_{ik}$ $(k=1,\ldots,n_{1i}) \sim F_{iY}(y)$, a subject-specific cumulative distribution function (cdf), where $F_{iY}(y)=P(Y_{ik}\leq y)$. Then, the subject-specific quantile function is defined as $Q_{iY}(p)= \inf\{y: F_{iY}(y)\geq p\}, p \in [0,1]$. The subject-specific quantile function is non-decreasing completely characterizes the distribution of the individual observations. Given $Y_{ik}$ s, the empirical quantile function can be calculated based on linear interpolation of order statistics \citep{parzen2004quantile} and serves as an estimate of the latent subject specific quantile function $Q_{iY}(p)$ \citep{yang2020quantile,yang2020random}. Similarly, assuming the repeated observations $X_{il}$ s come from $F_{iX}(x)$, one can define $Q_{iX}(p)= \inf\{x: F_{iX}(x)\geq p\}, p \in [0,1]$, which can be estimated by the empirical quantile function based on $X_{il}$ s. In particular, for a sample $(X_1,X_2,\ldots, X_n)$, let $X_{(1)}\leq X_{(2)}\leq \ldots,\leq X_{(n)}$ be the corresponding order statistics. The empirical quantile function, for $p \in [\frac{1}{n+1},\frac{n}{n+1}]$, is then given by, 
\begin{equation}
    \tilde{Q}(p)=(1-w)X_{(\lfloor(n+1)p\rfloor)}+wX_{(\lfloor(n+1)p\rfloor+1)}, \label{estquantile}
\end{equation}
where $w$ is a weight satisfying $(n+1)p=\lfloor(n+1)p\rfloor+w$. Based on this formulation and repeated observations $Y_{ik},X_{il}$, we can obtain the subject specific quantile functions $\tilde{Q}_{iY}(p)$ (for HR) and $\tilde{Q}_{iX}(p)$ (for PA) which are estimators of the true quantile functions ${Q}_{iY}(p)$,${Q}_{iX}(p)$. The empirical quantile functions are consistent \citep{parzen2004quantile} and are suitable for distributional representations due to several attractive mathematical properties \citep{powley2013quantile,ghosal2023distributional}, without requiring any smoothing parameter selection as in density estimation.

\subsection{Distributional Outcome Regression via Quantile Functions}
\label{dosdr mod}
We assume that scalar covariates $(z_1,z_2,\ldots,z_q) \in [0,1]^q$ without any loss of generality (e.g., achievable by linear transformation).
We posit the following distributional regression model, associating the distributional response ${Q}_{iY}(p)$ (which is non-decreasing itself) to the scalar covariates $z_{ij}$, $j=1,2,\ldots,q$, and a distributional predictor ${Q}_{iX}(p)$. We will refer to this as a distributional outcome regression via quantile functions
(DORQF).
\begin{eqnarray}
    E(Q_{iY}(p)\mid z_{i1},z_{i2},\ldots,z_{iq},Q_{iX}(p))=\beta_0(p) +\sum_{j=1}^{q}z_{ij}\beta_j(p) +h(Q_{iX}(p)),\\
    \epsilon_i(p) = Q_{iY}(p)-E(Q_{iY}(p)\mid z_{i1},z_{i2},\ldots,z_{iq},Q_{iX}(p)),\\
     Q_{iY}(p)=\beta_0(p) +\sum_{j=1}^{q}z_{ij}\beta_j(p) +h(Q_{iX}(p))+\epsilon_i(p). \label{dodsr}
\end{eqnarray}
We assume (a) residual error process $\epsilon_i(p)$ is a mean zero stochastic process of bounded variation satisfying $E(\epsilon_{i}(p)\mid z_{i1},z_{i2},\ldots,z_{iq},Q_{iX}(p))=0$ and (b) an $\epsilon_i(p)$ exists such that $p\rightarrow\beta_0(p) +\sum_{j=1}^{q}z_{j}\beta_j(p) +h(Q_{X}(p))+\epsilon(p)$ is non-decreasing \citep{zhang2011functional}. Here $\beta_0(p)$ is a distributional intercept and  $\beta_j(p)$ s are the distributional effects of the scalar covariates $z_j$ at quantile level $p$. The unknown nonparametric function $h(\cdot)$ captures the additive effect of the distributional predictor $Q_{iX}(p)$. In general, we do not directly observe the true latent quantile functions $Q_{iX}(p),Q_{iY}(p)$, rather we assume that we have subject-specific observations $\mathcal{X}_i=\{X_{i1}=Q_{iX}(u_{i1}),X_{i2}=Q_{iX}(u_{i2}),\ldots, X_{in_{2i}}=Q_{iX}(u_{in_{2i}})\}$ and $\mathcal{Y}_i=\{Y_{i1}=Q_{iY}(v_{i1}),Y_{i2}=Q_{iY}(v_{i2}),\ldots, Y_{in_{2i}}=Q_{iY}(v_{in_{1i}})\}$, where $u_{il},v_{ik}$s independently follow $U(0,1)$ distribution. Hence the observed data is given by  $D_i=\{Y_{ik}, X_{il},\*z_i;k=1,\ldots,n_{1i};l=1,\ldots,n_{2i}\}$, for subject $i=1,\ldots,n$. Based on $D_i$ we can calculate $\tilde{Q}_{iY}(p)$ and $\tilde{Q}_{iX}(p)$, as illustrated in Section 2.1 and use them in the DORQF model as proxy for the true latent quantile functions.

We make the following flexible and interpretable assumptions on the coefficient functions $\beta_j(\cdot),\hspace{1 mm} j=0,1,\ldots,q$ and on $h(\cdot)$ which ensures the predicted value of the response quantile function $Q_{iY}(p)$ conditionally on the predictors, $E(Q_{iY}(p)\mid z_{i1},z_{i2},\ldots,z_{iq},Q_{iX}(p))$ is non-decreasing, thus ensuring that the predicted quantile function stays in the restricted space. 

\begin{theorem}
Let the following conditions hold in the model (\ref{dodsr}).
\begin{enumerate}
       \item The distributional intercept $\beta_0(p)$ is non-decreasing.
    \item Any additive combination of  $\beta_0(p)$  with distributional slopes $\beta_j(p)$ is non-decreasing, i.e., $\beta_0(p)+\sum_{k=1}^{r} \beta_{j_k}(p)$  is non-decreasing for any sub-sample $\{j_1,j_2,\ldots,j_r\}\subset \{1,2,\ldots,q\}$.
    \item $h(\cdot)$ is non-decreasing.
\end{enumerate}
Then $E(Q_{Y}(p)\mid z_1,z_2,\ldots,z_q,Q_X(p))$  is non-decreasing.
\label{thm:monotone}
\end{theorem}
Note that $E(Q_{Y}(p)\mid z_1,z_2,\ldots,z_q,Q_X(p))$ is the predicted quantile function under the squared Wasserstein loss function, which is same as the squared error loss for the quantile functions. The proof is illustrated in Appendix A of the Supplementary Material. Assumptions (1) and (2) are much weaker and flexible than the monotonicity conditions of the QFOSR model in \cite{yang2020random}, where each of the function coefficients $\beta_j(p)$s is required to be monotone, whereas, we only impose monotonicity on the sum of functional coefficients. This is not just a technical aspect but this flexibility is important from a practical perspective, as it allows for capturing possible non-monotone association between the distributional response and individual scalar predictors $z_j$'s while still maintaining the required monotonicity of the predicted response quantile function. Condition (3) matches with the monotonicity assumption of the distributional regression model in \cite{ghodrati2022distribution} and in the absence of any scalar predictors, essentially captures the optimal transport map between the two distributions, after adjusting for scalars of interest - thus, providing a model general inferential framework compared to that in \cite{ghodrati2022distribution}. Thus, the above DORQF model extends the previous inferential framework for distributional response on scalar and contains both the QFOSR model and the distribution-on-distribution regression model as its submodels. More succinctly, in absence of distributional predictor we have, $
    Q_{iY}(p)=\beta_0(p) +\sum_{j=1}^{q}z_{ij}\beta_j(p)+\epsilon_i(p), \label{qfosr}
$
which is a quantile-function-on-scalar regression (QFOSR) model ensuring monotononicity under conditions (1),(2). Similarly, in absence of any scalar covariates, we have a distribution-on-distribution regression model 
$
    Q_{iY}(p)=\beta_0(p) +h(Q_{iX}(p))+\epsilon_i(p) \label{dodr}
$, which is a bit more general than the one considered in \cite{ghodrati2022distribution}, including a transnational effect $\beta_0(p)$. As a technical note, in model (\ref{dodsr}) function $h(\cdot)$ is identifiable only up to an additive constant, and in particular, the estimable quantity is the additive effect $\beta_0(p)+h(q_x(p))$ for a fixed $Q_X(p)=q_x(p)$. We impose the restriction $h(0)=0$ in order for $h(\cdot)$ to be estimable (see Section 2.3). \\ 

\hspace*{- 8mm}
\textbf{Remark 1:}
The assumptions (1)-(3) in Theorem 1 are a set of sufficient conditions for the predicted quantile process to be increasing, while allowing non monotonic association. Conditions (1) and (2) are also necessary conditions in order to $E(Q_{Y}(p)\mid z_{1},z_{2},\ldots,z_{q},Q_{X}(p))$ to be non-decreasing for all possible values of $z_{1},z_{2},\ldots,z_{q},Q_{X}(p)$ (note $(z_1,z_2,\ldots,z_q) \in [0,1]^q$ and $h(0)=0$). Condition (3) is a sufficient condition and captures the optimal transport map after adjusting for scalar covariates of interest.

\subsection{Estimation in DORQF}
\label{DODSR:estimation}
We follow a shape constrained estimation approach \citep{ghosal2022shape} for estimating the distributional effects $\beta_j(p)$ and the nonparamatric function $h()$ which naturally incorporates the constraints (1)-(3) of Theorem 1 in the estimation step. The univariate coefficient functions $\beta_j(p)$ ($j=0,1,\ldots,p$)
are modelled in terms of univariate expansions of Bernstein basis polynomials as \begin{equation}
    \beta_j(p)=\sum_{k=0}^{N}\beta_{jk}b_k(p,N),\hspace{2mm} \textit{where}\hspace{2mm}b_k(p,N)={N \choose k}p^k(1-p)^{N-k}, \hspace{2mm} \textit{for}\hspace{2mm } 0\leq p\leq 1.
    \label{uni}
\end{equation}
The number of basis polynomials depends on the degree of the polynomial basis $N$ (which is assumed to be same for all $\beta_j(\cdot)$ for computational tractability in this paper). The Bernstein polynomials $b_k(p,N)\geq 0$ and $\sum_{k=0}^{N}b_k(p,N)=1$. \cite{wang2012shape} and \cite{ghosal2022shape} illustrate that various shape constraints e.g., monotonicity, convexity, etc. can be reduced to linear constraints on the basis coefficients of the form $\+A_{N}\bm\beta_{j}^{N}\geq \bm 0$, where $\bm\beta_{j}^{N}=(\beta_{j0},\beta_{j1},\ldots,\beta_{jN})^T$ and $\+A_{N}$ is the constraint matrix chosen in a way to guarantee a desired shape restriction. In particular, in our context of DORQF, we need to choose constraint matrices $\+A_{N}$ in such a way which jointly ensure conditions (1),(2) in Theorem 1 and thus guarantee a non-decreasing predicted value of the response quantile function.
The nonparametric function $h(\cdot)$ is modelled similarly using univariate Bernstein polynomial expansion as
\begin{equation}
    h(x)=\sum_{k=0}^{N}\theta_{k}b_k(x,N),\hspace{2mm} \textit{where}\hspace{2mm}b_k(x,N)={N \choose k}x^k(1-x)^{N-k}, \hspace{2mm} \textit{for}\hspace{2mm } 0\leq x\leq 1.
    \label{uni:h}
\end{equation}
Since the domain of $h(\cdot)$ modelled via Bernstein basis is $[0,1]$, the quantile functions of the distributional predictor $Q_X(p)$ are transformed to a $[0,1]$ scale using linear transformation of the observed predictors. We make the assumption here that the distributional predictors are bounded, which is reasonable in the applications we are interested in. Henceforth, we assume $Q_X(p)\in [0,1]$ without loss of generality. Further, note that, $b_0(x,N)=1$, and since $\beta_0(p)$ already contains this constant term in the DORQF model (\ref{dodsr}), including the constant basis while modelling $h(\cdot)$ will lead to model singularity. Hence we drop the constant basis (i.e. the first term) while modelling $h(\cdot)$. In particular, $h(Q_{iX}(p))$ is modelled as $
    h(Q_{iX}(p))=\sum_{k=1}^{N}\theta_{k}b_k(Q_{iX}(p),N)$. Note that this is equivalent to imposing the constraint $h(0)=\theta_0=0$.   
The non-decreasing condition in (3) of Theorem 1 can again be specified as a linear constraint on the basis coefficients of the form $\+R\bm\theta\geq \bm 0$, where $\bm\theta=(\theta_{1},\ldots,\theta_{N})^T$, and $\+R$ is the constraint matrix. The DORQF model (\ref{dodsr}) can be reformulated in terms of basis expansions as
\begin{eqnarray}
     Q_{iY}(p)&=&\sum_{k=0}^{N}\beta_{0k}b_k(p,N) +\sum_{j=1}^{q}z_{ij}\sum_{k=0}^{N}\beta_{jk}b_k(p,N) +\sum_{k=1}^{N}\theta_{k}b_k(Q_{iX}(p),N)+\epsilon_i(p). \label{dodsr:exp1}\\
      &=& \*b_N(p)^T\bm{\beta_0} +\sum_{j=1}^{q}\*Z_{ij}^T(p)\bm{\beta_j} +\*b_N(Q_{iX}(p))^T\bm\theta+\epsilon_i(p). \notag
     \end{eqnarray}     
Here $\bm{\beta_j}=(\beta_{j0},\beta_{j1},\ldots,\beta_{jN})^T$, $\*b_N(p)^T=(b_0(p,N),b_1(p,N),\ldots,b_N(p,N))$, $\*b_N(Q_{iX}(p))^T=(b_1(Q_{iX}(p),N), b_2(Q_{iX}(p),N),\ldots,b_N(Q_{iX}(p),N)) $ and $\*Z_{ij}^T(p)=z_{ij}*\*b_N(p)^T$. Suppose that the qunatile functions $Q_{iY}(p), Q_{iX}(p)$ are evaluated on a grid $\mathcal{P}=\{p_1,p_2,\ldots,p_m\} \subset[0,1]$. Denote the stacked value of the quantiles for $i$th subject as $\*Q_{iY}=(Q_{iY}(p_1),Q_{iY}(p_2),\ldots,Q_{iY}(p_m))^T$. The DORQF model in terms of Bernstein basis expansion (\ref{dodsr:exp1}) can be reformulated as 
\begin{eqnarray}
    \*Q_{iY}&=& \+B_0\bm{\beta_0} +\sum_{j=1}^{q}\+W_{ij} \bm{\beta_j} +
    \+S_i\bm{\theta}+\bm\epsilon_i, \label{dodsr:exp2}
     \end{eqnarray}    
where $\+B_0=(\*b_N(p_1),\*b_N(p_2),\ldots,\*b_N(p_m))^T$,$\+W_{ij}=(\*Z_{ij}(p_1),\*Z_{ij}(p_2),\ldots,\*Z_{ij}(p_m))^T$ and $\+S_i=(\*b_N(Q_{iX}(p_1)), \*b_N(Q_{iX}(p_2)),\ldots,\*b_N(Q_{iX}(p_m)))^T$ and $\bm\epsilon_i$ are the stacked residuals $\epsilon_{i}(p)$s. The parameters in the above model are the basis coefficients $\bm\psi=(\bm{\beta_0}^T,\bm{\beta_1}^T,\ldots,\bm{\beta_q}^T,\bm\theta^T)^T$. For estimation of the parameters, we use the following least square criterion, which reduces to a shape constrained optimization problem. Namely, we obtain the estimates $\hat{\bm\psi}$ by minimizing residual sum of squares as \begin{equation}
     \hat{\bm\psi}=\underset{\bm\psi}{\text{argmin}}\hspace{2 mm}  \sum_{i=1}^{n}||\*Q_{iY}- \+B_0\bm{\beta_0} -\sum_{j=1}^{q}\+W_{ij} \bm{\beta_j} -
    \+S_i\bm{\theta}||_2^{2} \hspace{4mm} \textit{s.t \hspace{ 4 mm} $\+D\bm\psi\geq \bm 0$}. \label{funopt}
\end{equation}
The universal constraint matrix $\+D$ on the basis coefficients is chosen to ensure the conditions (1),(2),(3) in Theorem 1. In Appendix B of the Supplementary Material, we illustrate examples how the constraint matrix is formed in practice. The above optimization problem (9) can be identified as a quadratic programming problem \citep{goldfarb1982dual,goldfarb1983numerically}. R package \texttt {restriktor} \citep{res} can be used for performing the above optimization. Our estimation ensures that the shape restrictions are enforced everywhere and hence the predicted quantile functions are nondecreasing in the whole domain $p\in [0,1]$ as opposed to fixed quantile levels or design points in \cite{ghodrati2022distribution}. 

The order of the Bernstein polynomial basis $N$ controls the smoothness of the coefficient functions $\beta_j(\cdot)$ and $h(\cdot)$. We follow a truncated basis approach \citep{Ramsay05functionaldata,fan2015functional}, by restricting the number of BP basis to ensure the resulting coefficient functions are smooth. The optimal order of the basis functions is chosen via $V$-fold cross-validation method  \citep{wang2012shape} using cross-validated residual sum of squares defined as, 
 $CVSSE= \sum_{v=1}^{V}\sum_{i=1}^{n_v}||\*Q_{iY,v}-\hat{\*Q}_{iY,v}^{-v}||_2^{2}.$ Here $\hat{\*Q}_{iY}^{-v}$ is the fitted quantile values of observation $i$ within the $v$ th fold obtained from the constrained optimization criterion (\ref{funopt}) and trained on the rest $(V-1)$
folds.

\subsection{Uncertainty Quantification and Joint Confidence Bands}
\label{uncert q}
To construct confidence intervals and joint confidence bands for the distributional coefficients, we use the result that the constrained estimator $\hat{\bm\psi}$ in (\ref{funopt}) is the projection of the corresponding unconstrained estimator \citep{ghosal2022shape} onto the restricted space: $\hat{\bm\psi}_{r}=\underset{\bm\psi \in \bm\Theta_R }{\text{argmin}}\hspace{2 mm} ||\bm\psi-\hat{\bm\psi}_{ur}||^{2}_{\hat{\*\Omega}}$, for a non-singular matrix $\hat{\*\Omega}$. The complete procedure for uncertainty quantification and obtaining joint confidence bands is illustrated in Appendix C, D of the Supplementary Material. Based on the joint confidence band, it is possible to directly test for the global distributional effects $\beta(p)$ (or $h(x)$). The p-value for the test
$H_{0} : \beta(p)=0 \hspace{3mm} \hbox {for all $p\in [0,1]$} \hbox{ versus } H_{1} : \beta(p)\neq 0 \hspace{3mm} \hbox {for at least one $p\in [0,1]$} ,$ could be obtained based on the  $100(1-\alpha)\%$ joint confidence band 
for $\beta(p)$. In particular, following \cite{sergazinov2022case}, the p-value for the test can be defined as the smallest level $\alpha$ for which at least one of the  $100(1-\alpha)\%$ confidence intervals around $\beta(p)$ ($p \in \mathcal{P}$) does not contain zero. Alternatively, a nonparametric bootstrap procedure for testing the global effects  of scalar and distributional predictors is illustrated in Appendix E of the Supplementary Material which could be useful for  finite sample sizes and non Gaussian error process.
\section{Simulation Studies}
\label{simul}
In this Section, we investigate the performance of the proposed estimation and testing method for DORQF via simulations. To this end, we consider the following data generating scenarios.
\subsection{Data Generating Scenarios}
\subsection*{Scenario A1: DORQF, Both distributional and scalar predictor}
We consider the DORQF model given by $Q_{iY}(p)=\beta_0(p) +z_{i1}\beta_1(p) +h(Q_{iX}(p))+\epsilon_i(p)$. The distributional effects are taken to be $\beta_0(p)=2+3p$, $\beta_1(p)=sin(\frac{\pi}{2}p)$ and $h(x)=(\frac{x}{10})^3$. The scalar predictor $z_{i1}$ is generated independently from  a $U(0,1)$ distribution. The distributional predictor $Q_{iX}(p)$ is generated as $Q_{iX}(p)=c_i Q_{N}(p,10,1)$, where $Q_{N}(p,10,1)$ denotes the pth quantile of a normal distribution $N(10,1)$ and $c_i\sim U(1,2)$. The residual error process $\epsilon(p)$ is generated as $\epsilon_i(p)=A_i\beta_0(p)$, where $A_i\sim U(-0.5,0.5)$, which is of bounded variation. The above specifications guarantee a non-decreasing quantile function $Q_{iY}(p)$.

Since we do not directly observe these quantile functions $Q_{iX}(p),Q_{iY}(p)$ in practice we assume we have the subject-specific observations $\mathcal{X}_i=\{x_{i1}=Q_{iX}(u_{i1}),x_{i2}=Q_{iX}(u_{i2}),\ldots, x_{iL_{i1}}=Q_{iX}(u_{iL_{i1}})\}$ and $\mathcal{Y}_i=\{y_{i1}=Q_{iY}(v_{i1}),y_{i2}=Q_{iY}(v_{i2}),\ldots, y_{iL_{i2}}=Q_{iY}(v_{iL_{i2}})\}$, where $u_i,v_j$ s are independently generated from $U(0,1)$ distribution. For simplicity, we assume that $L_{i1}=L_{i2}=L$ many subject specific observations are available for both the distributional outcome and the predictor. Based on the observations $\mathcal{X}_i,\mathcal{Y}_i$ the subject specific quantile functions $Q_{iX}(p)$ and $Q_{iY}(p)$ are estimated based on empirical quantiles as illustrated in equation (\ref{estquantile}) on a grid of $p$ values $\in [0,1]$.
We consider number of individual measurements $L=200,400$ and training sample size $n=200,300,400$ for this data generating scenario. The grid  $\mathcal{P}=\{p_1,p_2,..\ldots,p_m\} \subset [0,1]$ is taken to be a equi-spaced grid of length $m=100$ in $[0.005,0.995]$. A separate sample of size $n_{t}=100$ is used as a test set for each of the above cases. Additional simulation scenarios (Scenario A2) illustrating the performance of the proposed test and a DORQF model with only distributional predictor (Scenario B) are reported in Appendix F of the Supplementary Material. We consider 100 Monte-Carlo (M.C) replications from simulation scenarios A1 and B to assess the performance of the proposed estimation method. For scenario A2, 500 replicated datasets are used to assess type I error and power of the proposed testing method.

\subsection{Simulation Results}
\subsection*{Performance under scenario A1:}
We evaluate the performance of our proposed method in terms of integrated mean squared error (MSE), integrated squared Bias (Bias$^2$) and integrated variance (Var). For the distributional effect $\beta_1(p)$, these are defined as $MSE=\frac{1}{M} \sum_{j=1}^{M}\int_{0}^{1}\{\hat{\beta}_{1}^{j}(p)-\beta_1(p)\}^2dp$, Bias$^2=\int_{0}^{1}\{\hat{\bar{\beta}}_{1}(p)-\beta_1(p)\}^2dp$, $Var=\frac{1}{M} \sum_{j=1}^{M}\int_{0}^{1}\{\hat{\beta}_{1}^{j}(p)-\hat{\bar{\beta}}_{1}(p)\}^2dp$. Here $\hat{\beta}_{1}^{j}(p)$ is the estimate of $\beta_1(p)$ from the $j$th replicated dataset and $\hat{\bar{\beta}}_{1}(p)=\frac{1}{M} \sum_{j=1}^{M}\hat{\beta}_{1}^{j}(p)$ is the M.C average estimate based on the M replications. Table \ref{tab:my-table} reports the squared Bias, Variance and MSE of the estimates of $\beta_1(p)$ for all the cases considered in scenario A1. The average choice of $N$ (order of the Bernstein polynomial basis) picked by the cross-validation method is reported in supplementary Table S3. MSE as well as squared Bias and Variance  are found to decrease and be negligible as sample size $n$ and the number of measurements $L$ increase, illustrating satisfactory accuracy of the proposed estimator.
\begin{table}[ht]
\centering
\caption{Integrated squared bias, variance and mean square error  of estimated $\beta_1(p)$ over 100 Monte-Carlo replications, Scenario A1. }
\label{tab:my-table}
\begin{tabular}{lllllll}
\hline
Sample Size              & \multicolumn{3}{l|}{L=200}                                                       & \multicolumn{3}{l}{L=400}                                                       \\ \hline
$\beta_1(p)$ & \multicolumn{1}{l}{Bias$^2$}              & \multicolumn{1}{l}{Var}    & MSE    & \multicolumn{1}{l}{Bias$^2$}              & \multicolumn{1}{l}{Var}    & MSE    \\ \hline
n= 200                   & \multicolumn{1}{l}{0.0020}               & \multicolumn{1}{l}{0.0508} & 0.0528 & \multicolumn{1}{l}{0.0017} & \multicolumn{1}{l}{0.0527} & 0.0544 \\ \hline
n= 300                   & \multicolumn{1}{l}{$0.0010$}               & \multicolumn{1}{l}{0.0439} & 0.0449 & \multicolumn{1}{l}{$0.0004$} & \multicolumn{1}{l}{0.0436} & 0.0440 \\ \hline
n= 400                   & \multicolumn{1}{l}{$3.9 \times 10^{-5}$} & \multicolumn{1}{l}{0.0391} & 0.0392 & \multicolumn{1}{l}{$3.9 \times 10^{-5}$} & \multicolumn{1}{l}{0.0372} & 0.0372 \\ \hline
\end{tabular}
\end{table}
Since, $h(x)$ is not directly estimable in the DORQF model, we consider estimation of the estimable additive effect $\gamma(p)=\beta_0(p)+h(q_x(p))$ at $q_x(p)= \frac{1}{n}\sum_{i=1}^{n}Q_{iX}(p)$. The performance of the estimates in terms of squared Bias, variance and MSE are reported in Supplementary Table S1, which again illustrates satisfactory performance of the proposed method in capturing the distributional effect of the distributional predictor $Q_{X}(p)$.

\begin{figure}[ht]
\begin{center}
\begin{tabular}{ll}
\includegraphics[width=.48\linewidth , height=.45\linewidth]{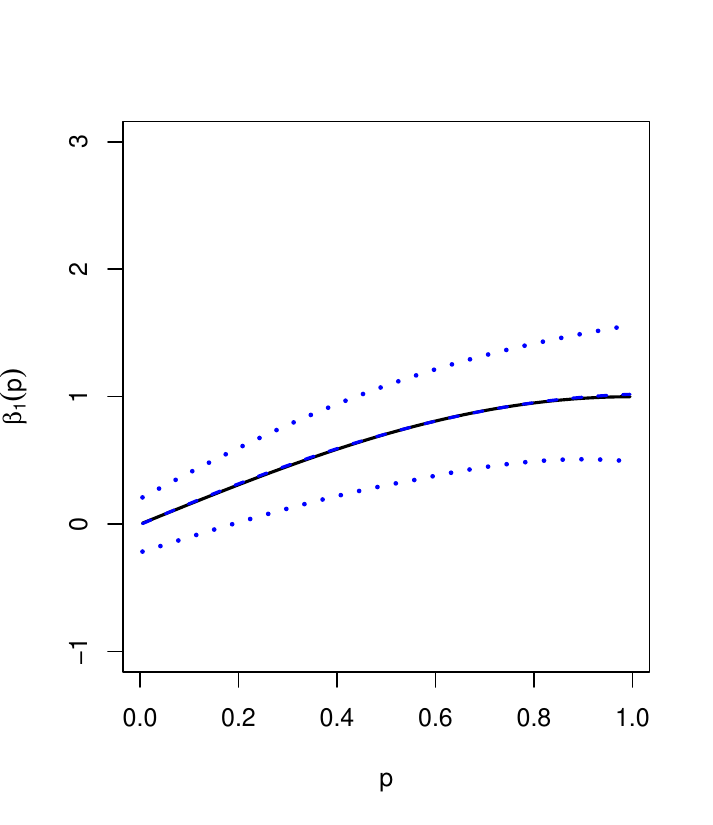} &
\includegraphics[width=.48\linewidth , height=.45\linewidth]{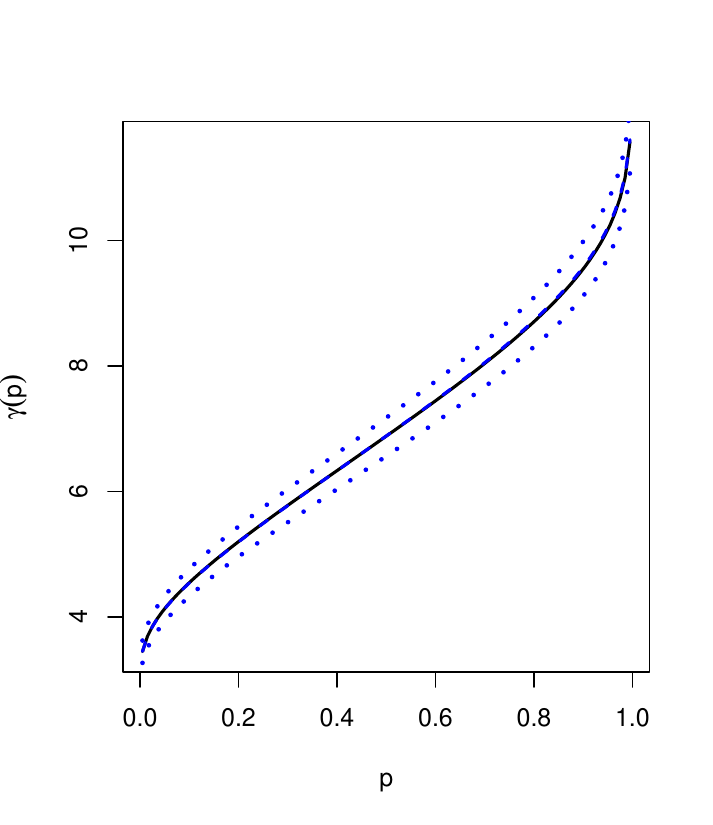}
\end{tabular}
\end{center}
\caption{Left: True distributional effect $\beta_1(p)$ (solid) and estimated $\hat{\beta_1}(p)$ averaged over 100 M.C replications (dashed) along with point-wise $95\%$ confidence interval (dotted), scenario A1, $n=400,L=400$. Right:  Additive effect $\gamma(p)=\beta_0(p)+h(q_x(p))$ (solid) at $q_x(p)= \frac{1}{n}\sum_{i=1}^{n}Q_{iX}(p)$ and its estimate $\hat{\gamma}(p)$ averaged over 100 M.C replications (dashed) along with point-wise $95\%$ confidence interval (dotted).}
\label{fig:fig2s1}
\end{figure}

The estimated M.C mean for the distributional effect $\beta_1(p)$ and $\gamma(p)$ along with their respective $95\%$ point-wise confidence intervals are displayed in Figure \ref{fig:fig2s1}, for the case $n=400, L=400$. The M.C mean estimates are superimposed on the true curves and along with the narrow confidence intervals, they illustrate low variability and high accuracy of the estimates.

As a measure of out-of- sample prediction performance, we report the average Wasserstein distance between the true quantile functions and the predicted ones in the test set defined as $WD=\frac{1}{n_t}\sum_{i=1}^{n_t}[\int_{0}^{1}\{Q_i^{test}(p)-\hat{Q_i}^{test}(p)\}^2 dp]^{\frac{1}{2}}$ in Supplementary Table S2. The low values of the average $WD$ metric and their M.C standard error indicate a satisfactory prediction performance of the proposed method. The performance of the proposed projection based joint confidence bands for $\beta_1(p)$ is investigated in Supplementary Table S3 which reports the coverage and width of the joint confidence bands for $\beta_1(p)$ for various choices of $N$ and for the case $L=200$. It is observed that the nominal coverage of $0.95$ lies within the two standard error limit of the estimated coverage in the all the cases, particularly for choices of $N$ picked by the proposed cross-validation method. 

The performance of the proposed test and estimation method for the additional simulation scenarios are reported in Appendix F of the supplementary material, where a similar impressive performance of the DORQF could be observed.

\section{Baltimore Longitudinal Study of Aging}
In this section, we apply DORQF to continuously monitored heart rate and physical activity data collected in Baltimore Longitudinal Study of Aging (BLSA). The BLSA study, initiated in 1958 by the National Institute of Aging Intramural Research Program \citep{schrack2018using}, is a comprehensive study of normative human aging. It consists of a cohort of community-dwelling volunteers who undergo rigorous health and functional assessments and are devoid of major chronic conditions upon enrollment. These participants are monitored throughout their lives, with testing intervals ranging from 1 to 4 years based on age. The sample of the current study constitutes of $n=781$  participants of age 50-97 years old who were recruited between September 2007 and November 2015. Supplementary Table S5 presents descriptive statistics of the sample. The Actiheart component of BLSA recorded minute-level heart rate (HR) and physical activity (PA) data for these older adults. Using Actiheart, a chest-worn heart rate and uniaxial activity monitor, participants were continuously monitored for 24 hours a day over 7-9 day period right after their on-site visit. Our main interest is in modelling and understanding age-related changes in daily life heart rate reserve, while accounting for gender and BMI as well as for the daily composition of physical activity. For the analysis in the current paper, we consider the HR and PA data collected on all the days for each BLSA participant during their first BLSA visit. The average number of days of data for each participant was 8. We focus on 8AM-8PM as the main period of wake-time activity and calculate distributional representations of both HR and PA data via corresponsing subject-specific quantile functions, $Q_{iY}(p)$ for minute-level heart rate and $Q_{iX}(p)$ for minute-level log-transformed activity counts. Supplementary Figure S4 displays these subject-specific quantile functions of HR and PA for all the study participants. 

We start with a simple linear regression to explore associations between daily average of heart rate and age, gender (Male=1, Female=0), BMI and daily average of ACs
$$\mu_{H,i}= \theta_0+ \theta_1 age_i +\theta_2 gender_i+ \theta_3 BMI_i +\theta_4 \mu_{A,i}+\epsilon_i,$$
where $\mu_{H,i},\mu_{A,i}$ are the subject specific averages of daily heart rate and daily activity counts. Supplementary Table S5 reports the results of this model. Average daily heart rate is found to be negatively associated with age, significantly lower for males, positively associated with average daily activity, and not significantly associated with BMI. The above results may be useful as a starting point to establish directions of effects.

 Next, we will try to get a much more detailed picture by applying the proposed DORQF model as follows:
\begin{equation}
    Q_{iY}(p)=\beta_0(p) +age_{i}\beta_{age}(p)+ BMI_{i}\beta_{BMI}(p)+gender_{i}\beta_{gender}(p) +h(Q_{iX}(p))+\epsilon_i(p) \label{dodsr:real}.
\end{equation}
Scalar predictors including age and BMI  are transformed to be in $[0,1]$ using monotone linear transformations (for example, $age_{[0,1]} = \frac{age-age_{min}}{age_{max}-age_{min}}$). Quantile function predictors are normalized to [0,1] using similar transformation on AC (log-transformed) ($AC_{[0,1]} = \frac{AC-AC_{min}}{AC_{max}-AC_{min}}$), where $AC_{min},AC_{max}$ are calculated based on all AC observations across all participants. The common degree of the Bernstein polynomial basis used to model all the distributional coefficient was chosen via five-fold cross-validation method that resulted in $N=5$. 

\begin{figure}[ht]
\centering
\includegraphics[width=1\linewidth , height=0.6\linewidth]{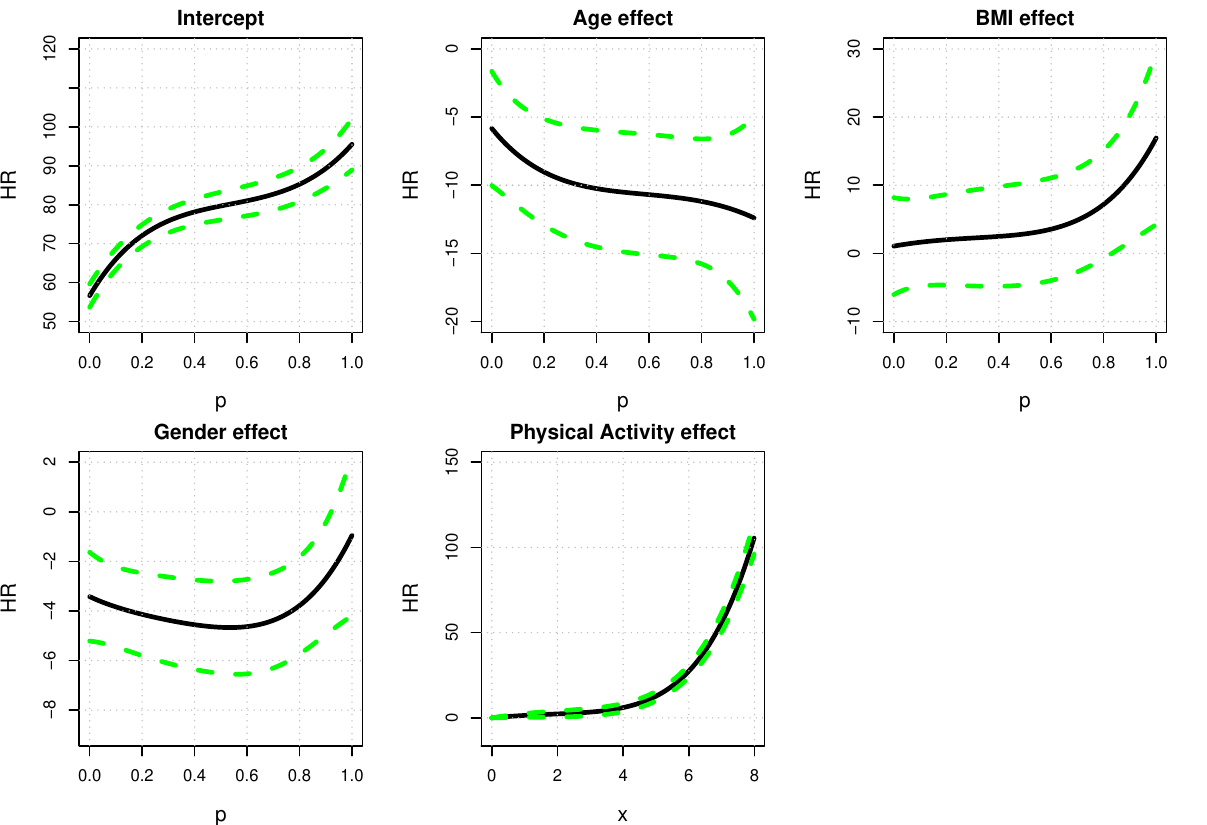}
\caption{Estimated distributional effects (solid black) along with their joint 95$\%$ confidence bands (dashed green) for intercept ($\beta_0(p)$), age ($\beta_{age}(p)$), BMI ($\beta_{BMI}(p)$) and gender (Male = 1, $\beta_{gender}(p)$) on heart rate along with the estimated link function $h(\cdot)$ (bottom middle panel) between the distributions of heart rate and physical activity.}
\label{fig:fig5}
\end{figure}

\begin{figure}[ht]
\centering
\includegraphics[width=1\linewidth , height=0.6\linewidth]{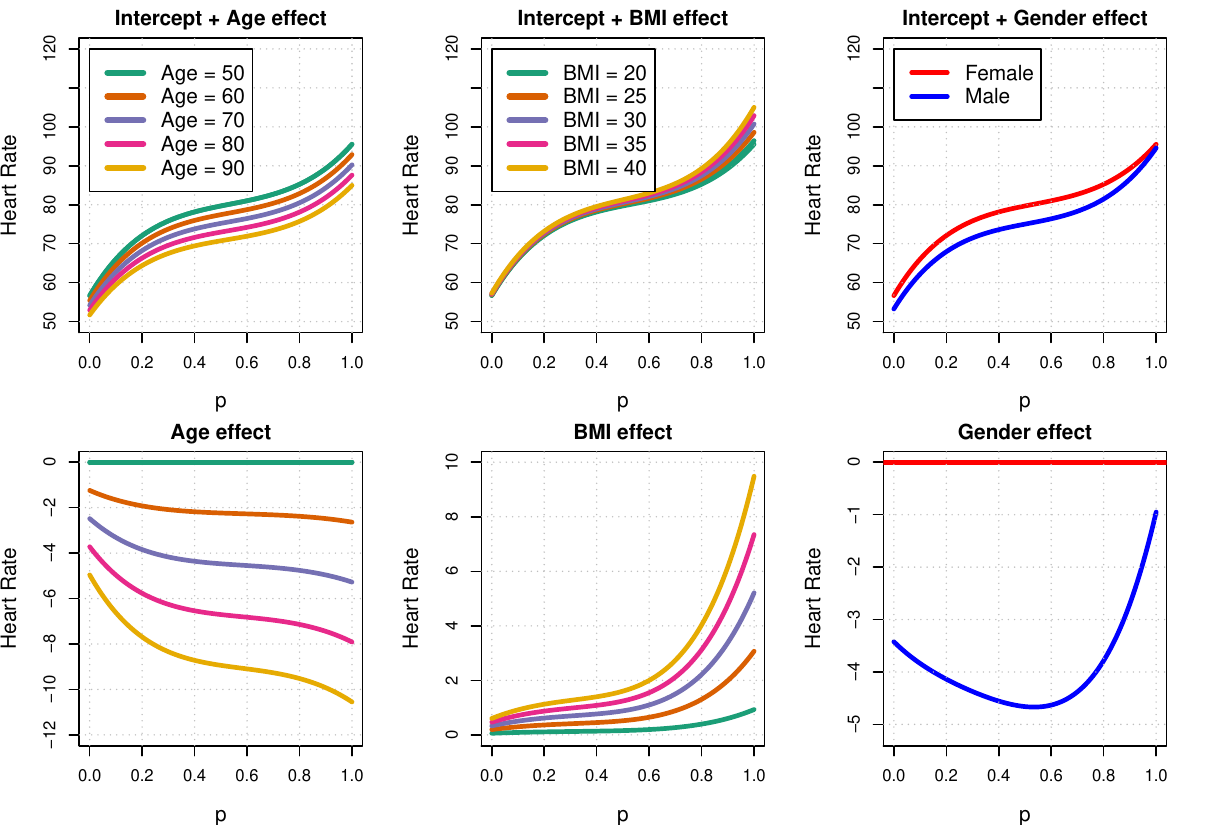}
\caption{First, second, and third columns show the combined with intercept (top) and individual effects of age, BMI, and gender, respectively, for corresponding representative ranges for each variable.}
\label{fig:fig4}
\end{figure}

Figure \ref{fig:fig5} shows estimated distributional effects along with their asymptotic 95$\%$ joint confidence bands. Figure \ref{fig:fig4} illustrates the combined with intercept and individual effects of age, BMI, and gender.
The p-values from the joint confidence band-based global test for the intercept and age, BMI, gender, and distribution of activity counts are  $1\times10^{-6},1\times10^{-6},0.004,9\times10^{-4}$ and $1\times10^{-6}$, respectively, resulting in the significance of all the predictors. The p-values from the nonparametric bootstrap test for age, BMI, gender, and distribution of activity counts are calculated to be $\leq 0.01$, $0.04$, $\leq 0.01$, $\leq 0.01$ respectively (based on $B=100$ bootstraps), leading to similar conclusions.

The estimated distributional intercept $\hat{\beta}_0(p)$ is monotone (increasing) and represents a reference to be used while interpreting the effect of predictors. Based on joint confidence bands, the estimated effect of age is significant across all quantile levels. The effect is negative and appears to be decreasing with $p$. Thus, moderate-to-high levels of daily life heart rate reserve decrease with age at a much faster (approximately doubled) rate compared to resting-to-light levels. 


The global effect of BMI is found to be significantly associated with DL-HRR. Joint confidence bands are only above zero for a very large values of $p$ ($p > 0.9$), where BMI exhibits highly nonlinear effect. Note that BMI was not significant in the simple scalar regression, which highlights that DORQF can uncover local with respect to $p$ effects. 

The estimated effect of gender illustrates that females have higher heart rate \citep{antelmi2004influence,prabhavathi2014role} compared to males across all quantile levels after adjusting for age, BMI and distributional composition of daily PA. 
The lower heart rate in males compared to females can be attributed to size of the heart, which is typically smaller in females than males \citep{prabhavathi2014role} and thus need to beat faster to provide the same output. 

The estimated monotone map (under the constraint $h(0)=0$) between DL-HRR and daily composition of PA $\hat{h}(x)$ is found to be highly nonlinear and convex, illustrating a non-linear dependence of levels of heart rate on the corresponding levels of physical activity. The convex nature of the map points out to the accelerated increase in heart rate quantiles with an increase in the corresponding quantile levels of PA \citep{leary2002morning}. To additionally demonstrate the effect of daily life PA on DL-HRR, 
Supplementary Figure S7 shows the predicted heart rate quantile functions as a function of sequential deviations ($\delta=0,0.5,0.10,0.15$ for scaled PA in $[0,1]$) from the barycenter of the PA distribution ($\bar{q}_x(p)= \frac{1}{n}\sum_{i=1}^{n}Q_{iX}(p)$) while holding age (=65), gender (=female) and BMI (=25) fixed.

Decreasing distributional effect of age and non-monotone distributional effect of gender  illustrate serious limitations of enforcing individual (non-decreasing) monotonicity in the model of \cite{yang2020random}. This clearly demonstrates that DORQF method is more flexible in requiring only joint monotonicity of the entire regression structure. In additional analysis, we tested for possible interaction between gender and age, and between gender and BMI. The p-values from the joint confidence band-based global test were $0.495$ and $0.06$ respectively, indicating no significant gender difference (at 5\% significance level) in the effect of aging and BMI on the heart rate distribution.

\begin{figure}[ht]
\begin{center}
\begin{tabular}{ll}
\includegraphics[width=.5\linewidth , height=.5\linewidth]{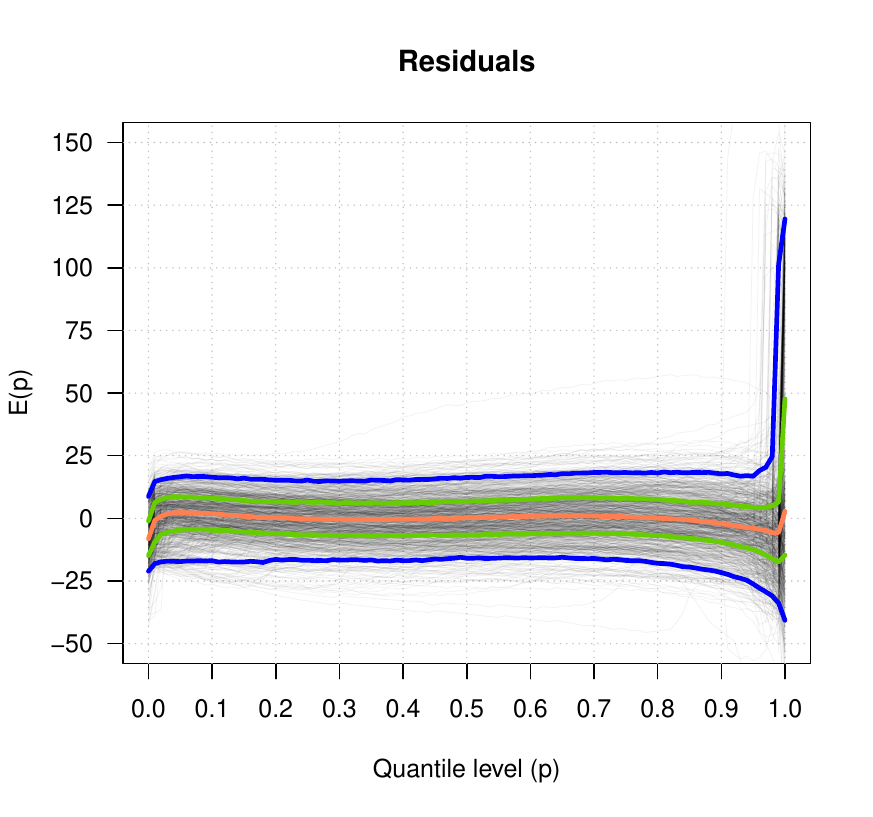} &
\includegraphics[width=.5\linewidth , height=.5\linewidth]{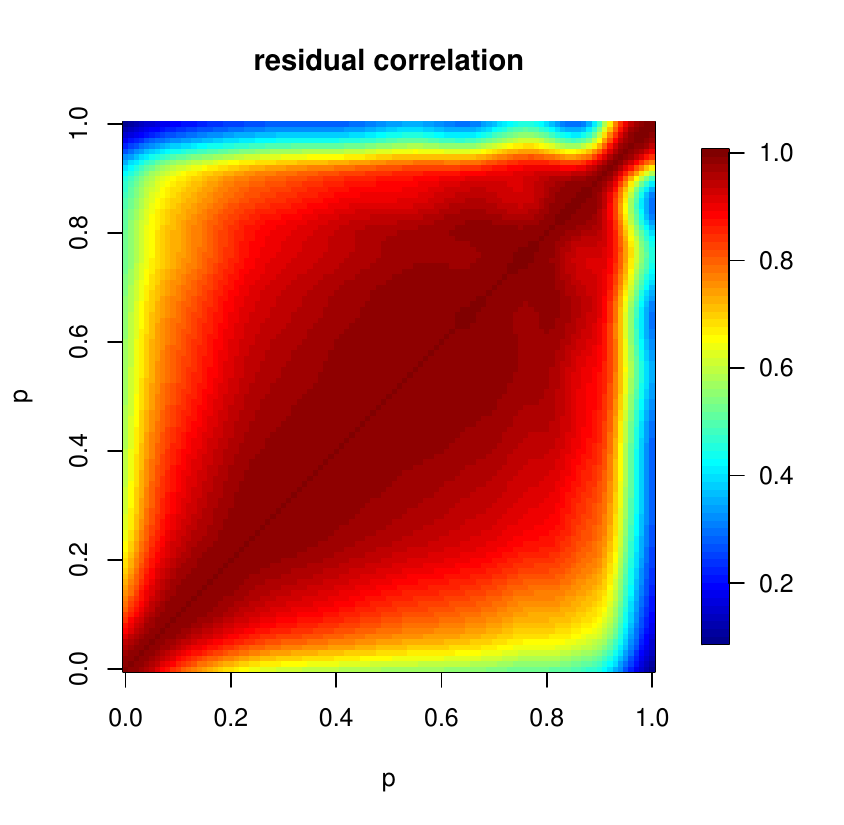}
\end{tabular}
\end{center}
\caption{Left: Residuals (black curves) from DORQF model (14) fitted in BLSA data with corresponding pointwise percentile curves (median (red), 25\% and 75\% (green), and 5\% and 75\% (blue) curves). Right: Estimated correlation surface of the residuals.}
\label{fig:figres}
\end{figure}

We explore the fitted residuals from the DORQF model (14). The fitted residuals along with it's correlation surface are displayed in Figure \ref{fig:figres}.
The fitted residuals are not necessarily non-decreasing. Across almost all quantile levels, $p$, they fluctuate within $[-20,20]$ beats per minute range. Substantially higher variability and a few outlying values can be noticed in the maximal quantile levels ($p>0.95$). There appears to be a strong non-stationary correlation within the residual process at different quantiles $p$, $p^{'}$, which gradually decreases as the distance $|p-p^{'}|$ increase. The leading eigenfunctions (explaining $\geq99\%$) of the corresponding covariance surface are shown in Supplementary Figure S8. The first functional principal component (FPC) can be noticed to capture a weighted average of the residual process across quantiles, while the higher order FPCs capture contrasts across different quantiles. To check the strict exogeneity assumption of the residual process, we plot the FPC scores of the fitted residuals against age, BMI and gender \citep{chiou2007diagnostics}. These are shown in Supplementary Figures S9- S11. The residuals appear to be uncorrelated with the scalar predictors. The in-sample predictive performance of the proposed DORQF method in terms of root-mean-square prediction error (RMSPE) between the observed and predicted quantile functions is shown in Supplementary Figure S12 across different quantiles. The prediction error is observed to be considerably higher across upper percentiles ($p > 0.95$), and might be attributed  to sparsity of heart-rate data at the right tail, and possible presence of a few outliers with extremely high maximal heart-rate.

Finally, we compare the predictive performance of DORQF with that of the distribution-on-distribution regression model by \cite{ghodrati2022distribution} based on isotonic regression (referred as DODR-ISO). Supplementary Figure S13 displays the leave-one-out-cross-validated (LOOCV) predicted quantile functions of heart rate from the DORQF and the DODR-ISO methods. Additionally, we also fit an unconstrained (standard) functional regression model corresponding to equation (10) using the \texttt{"pffr"} function within the \texttt{refund} package in R \citep{Rsoft}. We define the measure LOOCV $R$-Squared as $R_{loocv}^2=1- \frac{\sum_{i=1}^{N}\int_{0}^{1}\{Q_i(p)-\hat{Q_i}^{loocv}(p)\}^2 dp}{\sum_{i=1}^{N}\int_{0}^{1}\{Q_i(p)-\bar{Q}\}^2 dp},$ where $\bar{Q}=\frac{1}{N}\sum_{i=1}^{N}\int_{0}^{1}Q_i(p)dp$ to compare the out-of-sample prediction accuracy of these three methods. The $R_{loocv}^2$ value for the DORQF, DODR-ISO and the unconstrained functional model are calculated to be $0.59$, $0.47$, $0.49$ respectively. This clearly illustrates that the proposed DORQF model is able to predict daily life heart rate reserve more accurately with the use of additional scalar predictors including age, gender, and BMI. The performance gain of DORQF is quite substantial ($20\%$) compared to the unconstrained functional model, illustrating the usefulness of the DORQF model beyond technical innovation. For the unconstrained functional model the estimated distributional coefficients and its bootstrapped $95\%$ point-wise confidence intervals are shown in Supplementary Figure S5, those for DOQRF are shown in Supplementary Figure S6 with projection-based point-wise confidence intervals. Although the estimated distributional coefficients largely appear to be similar, we note that the proposed DORQF based confidence intervals are able to capture the significant effect of BMI on maximal HR, unlike the unconstrained approach.
The average width of the point-wise confidence intervals are reported in Supplementary Table S7. The average width of the point-wise confidence intervals are found to be lower for all the estimated coefficients from the DORQF method further highlighting its advantages.

\section{Discussion}
In this article, we have developed a flexible distributional outcome regression via quantile functions that can handle both scalar and distributional predictors. A novel functional regression structure was developed to guarantee outcome monotonicity under minimally restrictive constraints that provides more flexible modelling framework compared to existing frameworks. Inferential tools are developed that include projection-based asymptotic joint confidence bands and a global test of statistical significance for estimated functional regression coefficients. Numerical analysis using simulations illustrate an accurate performance of the estimation method. The proposed test is also shown to maintain the nominal test size and have a satisfactory power. In addition, a bootstrap test is proposed that could be particularly useful in finite sample sizes.

DORQF has been applied to BLSA Actiheart data to model age-related changes in daily life heart rate reserve while accounting for gender, body composition, and daily distribution of physical activity. The major finding is that highest levels of daily life heart rate reserve decrease with age at approximately twice faster rate compared to lower levels of DL heart rate reserve. To the best of our knowledge, this is a novel finding that has not been previously reported in literature that typically assumes a relatively constant age-related rates of declines across main levels of HR reserve \citep{tanaka2001age}. This demonstration of flexibility of DORQF is especially encouraging in view recent large nation-wide studies including Adolescent Brain Cognitive Development \citep{karcher2021abcd} and All of US \citep{all2019all} that integrated heart rate and activity Fitbit data in their data collection. Such major NIH-funded studies can readily benefit from the proposed framework to better characterize and quantify the associations between daily life heart rate and physical activity data and human development, health and disease. Beyond a large number of possible applications for DORQF in clinical and epidemiological studies collecting distributional observations, DORQF can be used for estimation of treatment effects in primary or secondary endpoints quantified via distributions.

There are multiple research directions that remain to be explored based on this work. In developing our method, we have implicitly assumed that there are enough measurements available per subject to accurately estimate quantile functions. Scenarios with only a few sparse measurements pose a practical challenge and will need careful handling. 
Our quantile based approach synchronize subjects across quantile levels. From one perspective, this could be seen as an advantage of the  proposed DORQF method as the quantile functions of heart rate $Q_{iY}(p)$ (outcome) all have the same domain $p \in (0,1)$ as opposed to in a density (or distribution) based method, where this would be modelled as $f_i(x)$, and the support of $x \in \mathcal{S}_i$ could have been very different for different subjects in extreme cases. Quantile synchronization helps in estimation of the distributional coefficients of the scalar predictors $\beta_j(p)$, which have the same uniform domain $p \in [0,1]$. For the distributional predictor (physical activity), as long as they are bounded in an interval, we scale them to $[0,1]$, which is required for estimation step of $h(\cdot)$. However, markedly different profiles of $Q_{iX}(p)$s might affect the estimation of $h (\cdot)$, especially, at the boundaries due to sparsity of observations.
Other study designs such as adapting to multilevel \citep{goldsmith2015generalized} data and longitudinal designs \citep{park2015longitudinal} would be also interesting to explore. The DORQF method could be extended to a temporally varying distributional framework, for modelling temporal evolution of subject-specific distributions of data \citep{ghosal2021scalar}, and the proposed framework is a significant and necessary first step in that direction. Another interesting direction of research could be to extend these models beyond the additive paradigm, for example,the single index model \citep{jiang2011functional}. Extending the proposed method to such more general and complex models would be computationally challenging, nonetheless merits future attention because of potentially diverse applications. 

\section*{Supplementary Material}
Appendix A-F along with the Supplementary Tables and Supplementary Figures referenced in this article are available online as Supplementary Material.

\section*{Software}
Software implementation via R \citep{Rsoft} and illustration of the proposed framework is available with this article and will be made publicly available on Github.

\section*{Acknowledgement}
The data presented in this article was obtained from \url{https://www.blsa.nih.gov} and the analysis plan was approved by BLSA. The BLSA is supported by the Intramural Research Program, NIA/NIH. The authors thank the staff and participants of the BLSA for their important contributions.
\vspace*{-5 mm}
\bibliographystyle{Chicago}
\bibliography{refs}
\end{document}


\def\spacingset#1{\renewcommand{\baselinestretch}%
{#1}\small\normalsize} \spacingset{1}


\if0\blind
{
  \title{\bf Supplementary Material for Distributional outcome regression via quantile functions and its application to modelling continuously monitored heart rate and physical activity}
 \author{Rahul Ghosal$^{1,\ast}$, Sujit Ghosh$^{2}$, Jennifer A. Schrack$^{3}$, Vadim Zipunnikov$^{4}$ \\
  \\
$^{1}$ Department of Epidemiology and Biostatistics, University of South Carolina \\
$^{2}$Department of Statistics, North Carolina State University\\
$^{3}$ Department of Epidemiology,  Johns Hopkins Bloomberg \\School of Public Health\\
$^{4}$ Department of Biostatistics, Johns Hopkins Bloomberg \\School of Public Health 
}
  \maketitle
} \fi

\if1\blind
{
  \bigskip
  \bigskip
  \bigskip
  \begin{center}
    {\LARGE\bf Supplementary Material for Distributional outcome regression via quantile functions and its application to modelling continuously monitored heart rate and physical activity}
\end{center}
  \medskip
} \fi

\bigskip

\vfill

\newpage
\spacingset{1.5} 

\section{Appendix A: Proof of Theorem 1}
The predicted outcome quantile function is the conditional expectation of the outcome quantile function based on the distribution outcome regression via quantile functions (DORQF) model (2) and is given by, 
\begin{equation}
    E(Q_{Y}(p)\mid z_1,z_2,\ldots,z_q,Q_X(p))=\beta_0(p) +\sum_{j=1}^{q}z_{j}\beta_j(p) +h(Q_{X}(p)). \label{dodsr:cond}
\end{equation}
We will show conditions (1)-(3) are sufficient conditions to ensure $E(Q_{Y}(p)\mid z_1,z_2,\ldots,z_q,Q_X(p))$
is non-decreasing. Let us assume $0\leq z_j \leq 1, \forall j= 1,2,\ldots, J$, without loss of generality. It is enough to show $T_1(p)=\beta_0(p) +\sum_{j=1}^{q}z_{j}\beta_j(p)$ and $T_2(p)=h(Q_{X}(p))$ both are non decreasing. The second part is immediate as both $Q_X(\cdot)$ and $h(\cdot)$ (by condition (3)) are non decreasing. To complete the proof we only need to show $T_1(p)$ is non decreasing. 

$T_1^{\prime}(p)=\beta_0^{\prime}(p)+\sum_{j=1}^{q}z_j\beta_j^{\prime}(p)$. Enough to show $T_1^{\prime}(p)\geq 0$ for all $(z_1,z_2,\ldots,z_q) \in [0,1]^q$. Note that this is a linear function in $(z_1,z_2,\ldots,z_q) \in [0,1]^q$. By the well-known Bauer's principle the minimum is attained at the boundary points $B=\{(z_1,z_2,\ldots,z_q):z_j\in \{0,1\}\}$. Hence, the sufficient conditions are $\beta_0^{\prime}(p)\geq0$ and  $\beta_0^{\prime}(p)+\sum_{k=1}^{r} \beta_{j_k}^{\prime}(p) \geq0$ for any sub-sample $\{j_1,j_2,\ldots,j_r\}\subset \{1,2,\ldots,q\}$, which follows from condition (1) and (2).

\section{Appendix B: Examples of DORQF}
\textbf{Example 1: Single scalar covariate ($q=1$) and a distributional predictor}\\
We consider the case where there is a single scalar covariate $z_1$ ($q=1$) and a distribution predictor $Q_X(p)$. In this case, the DORQF model is given by 
$Q_{iY}(p)=\beta_0(p) +z_{i1}\beta_1(p) +h(Q_{iX}(p))+\epsilon_i(p).$ The sufficient conditions (1)-(3) for non-decreasing quantile functions in this case reduces to: A) The distributional intercept $\beta_0(p)$ is non-decreasing
    B) $\beta_0(p)+\beta_{1}(p)$  is non-decreasing
    C) $h(\cdot)$ is non-decreasing. Note that the above conditions do no enforce $\beta_{1}(p)$ to be non-decreasing. Once the coefficient functions are modelled in terms of Bernstein basis expansions, conditions (A)-(C) can be be enforced via the following linear restrictions on the basis coefficients i.e.,
$\^A_{N}\bm\beta_{0}\geq0, [\^A_{N} \hspace{2 mm}\^A_{N}] (\bm\beta_{0}^T,\bm\beta_{1}^T)^T \geq0, \^A_{N-1}\bm\theta\geq0$. 
Here $\^A_N$ is a constraint matrix which imposes monotonicity on functions $f_N(x)$ modelled with Bernstein polynomials as
$f_N(x)=\sum_{k=0}^{N}\beta_{k}b_k(x,N),\hspace{2mm} \textit{where}\hspace{2mm}b_k(x,N)={N \choose k}x^k(1-x)^{N-k}, \hspace{2mm} \textit{for}\hspace{2mm } 0\leq x\leq 1.$ The derivative is given by $f_N^{\prime}(x)=N\sum_{k=0}^{N-1}(\beta_{k+1}-\beta_{k})b_k(x,N-1).$ 
Hence if $\beta_{k+1}\geq\beta_{k}$ for $k=0,1,\ldots, N-1$, $f_{N}(x)$ is non decreasing, which is achieved with the constraint matrix $\^A_{N}$. The combined linear restrictions on the parameter $\bm\psi=(\bm{\beta_0}^T,\bm{\beta_1}^T,\bm\theta^T)^T$ is given by $\^D\bm\psi \geq0$. The matrices $\^A_{N},\^D$ are given in equation \ref{conmat}. 

\begin{equation}
 \^A_{N}\equiv \begin{pmatrix}
 -1 & 1&0  & \dots & 0\\
 0 & -1 &1 &0 & \dots\\
 & & \ddots & &\\
 0 &\dots &0& -1 & 1\\
 \end{pmatrix},
 \^D= \begin{pmatrix}
 \^A_N & 0 & 0\\
 \^A_N & \^A_N &0\\
 0   &  0  & \^A_{N-1}
 \end{pmatrix}. \label{conmat}
\end{equation}

\hspace*{- 8 mm}
\textbf{Example 2: Two scalar covariates ($q=2$) and a distributional predictor}\\
We illustrate the estimation for DORQF where there are two scalar covariates $z_1,z_2$ ($q=1$) and a single distribution predictor $Q_X(p)$. The DORQF model is given by
$Q_{iY}(p)=\beta_0(p) +z_{i1}\beta_1(p)+z_{i2}\beta_2(p) +h(Q_{iX}(p))+\epsilon_i(p).$ The sufficient conditions (1)-(3) of Theorem 1 in this case reduce to : A) The distributional intercept $\beta_0(p)$ is non-decreasing.
 B)  $\beta_0(p)+\beta_{1}(p)$, $\beta_0(p)+\beta_{2}(p)$, $\beta_0(p)+\beta_{1}(p)+\beta_{2}(p)$  is non-decreasing.
 C) $h(\cdot)$ is non-decreasing. Note that condition B) illustrates that as the number of scalar covariates increase we have more and more combinatorial combinations of the coefficint functions restricted to be non-decreasing. Similar to Example 1, Conditions (A)-(C) again become linear restrictions on the basis coefficients of the form $\^D\bm\psi \geq0$, where the constraint matrix is given by $\^D= \begin{pmatrix}
 \^A_N & 0 & 0 & 0\\
 \^A_N & \^A_N &0 &0\\
 \^A_N & 0 &\^A_N &0\\
 \^A_N & \^A_N &\^A_N &0\\
 0   &  0 &0 & \^A_{N-1}
 \end{pmatrix}.$\\
As the number of restrictions increase the parameter space becomes smaller and smaller, which can result in a faster convergence of the optimization algorithm.

\section{Appendix C: Uncertainty Quantification and Joint Confidence Bands}
\label{uncert q}
To construct confidence intervals, we use the result that the constrained estimator $\hat{\bm\psi}$ in equation (9) of the paper is the projection of the corresponding unconstrained estimator \citep{ghosal2022shape} onto the restricted space: $\hat{\bm\psi}_{r}=\underset{\bm\psi \in \bm\Theta_R }{\text{argmin}}\hspace{2 mm} ||\bm\psi-\hat{\bm\psi}_{ur}||^{2}_{\hat{\*\Omega}}$, for a non-singular matrix $\hat{\*\Omega}$. The restricted parameter space is given by $ \bm\Theta_R=\{\bm\psi \in R^{K_n}: \^D\bm\psi\geq \bm 0  \}$. The DORQF model (8) can be reformulated as 
$\*Q_{iY}=\^T_{i} \bm{\psi} +\bm\epsilon_i$, 
where $\^T_i=[\^B_0 \hspace{2mm}\^W_{i1} \hspace{2mm}\^W_{i2},\ldots,\^W_{iq} \hspace{2mm} \^S_i]$ . The unrestricted and restricted estimators are given by,
$\hat{\bm\psi}_{ur}=\underset{\bm\psi \in R^{K_n} }{\text{argmin}}\hspace{2 mm}  \sum_{i=1}^{n}||{\*Q_{iY}}-\^T_i\bm\psi ||_2^{2} \hspace{4mm}$ and $
      \hat{\bm\psi}_{r}=\underset{\bm\psi \in \bm\Theta_R }{\text{argmin}}\hspace{2 mm}  \sum_{i=1}^{n}||{ \*Q_{iY}}-\^T_i\bm\psi ||_2^{2} \hspace{4mm}$. Let us denote $\*Q_{Y}^{T}=( \*Q_{1Y}, \*Q_{2Y},\ldots, \*Q_{nY})^T$ and $\^T=[\^T_1^{T},\^T_2^{T},\ldots,\^T_n^{T}]^T$. Then we can write,
$$\frac{1}{n}||\*Q_{Y}-\^T\bm\psi||_2^{2}=\frac{1}{n}||\*Q_{Y}-\^T\hat{\bm\psi}_{ur}||_2^{2}+\frac{1}{n}||\^T\hat{\bm\psi}_{ur}-\^T{\bm\psi}||_2^{2}.$$
Hence $\hat{\bm\psi}_{r}=\underset{\bm\psi \in \bm\Theta_R }{\text{argmin}}\hspace{2 mm} ||\bm\psi-\hat{\bm\psi}_{ur}||^{2}_{\hat{\*\Omega}},$ where
$\hat{\bm\Omega}=\frac{1}{n}\sum_{i=1}^{n}\^T_i^{T}\^T_i^{}$ and $\bm\Omega=E(\hat{\bm\Omega})$ is non-singular. Thus, we can use the projection of the large sample distribution of $\sqrt{n}(\hat{\bm\psi}_{ur}-\bm\psi^0)$ to approximate the distribution of 
$\sqrt{n}(\hat{\bm\psi}_{r}-\bm\psi^0)$. Now $\sqrt{n}(\hat{\bm\psi}_{ur}-\bm\psi^0)$ is asymptotically distributed as $N(0,\bm\Delta)$ under suitable regularity conditions \citep{huang2004polynomial,huang2002varying} for general choice of basis functions (holds true for finite sample sizes if $\epsilon(p)$ is Gaussian), where $\bm\Delta$ can be estimated by a consistent estimator. In particular, we use a sandwich covariance estimator corresponding to model $\*Q_{iY}=\^T_{i} \bm{\psi} +\bm\epsilon_i$, for estimating $\bm\Delta$ following a functional principal component analysis (FPCA) approach \citep{ghosal2022score} for estimation of the covariance matrix of the residuals $\bm\epsilon_i$ ($i=1,\ldots,n$). Details of this estimation procedure is included in Appendix D of the Supplementary Material.  

Let us consider the scenario with a single scalar covariate and distributional predictor for simplicity of illustration. The Bernstein polynomial approximation of $\beta_1(p)$ be given by $\beta_{1N}(p)=\sum_{k=0}^{N}\beta_{k}b_{1k}(p,N)=\rho_{K_n}(p)^{'}\bm\beta_1$. Algorithm 1 (presented below) is used to obtain an asymptotic $100(1-\alpha)\%$ joint confidence band for the true coefficient function $\beta_1^0(p)$, corresponding to a scalar predictor of interest. Here $\beta_1^0(p)$ denotes the true distributional coefficient $\beta_1(p)$. The algorithm relies on two steps i) Use the asymptotic distribution of $\sqrt{n}(\hat{\bm\psi}_{r}-\bm\psi^0)$ to generate samples from the asymptotic distribution of  $\hat{\beta}_{1r}(p)$ (these can be used to get point-wise confidence intervals)  ii) Use the generated samples and the supremum test statistic \citep{meyer2015bayesian,cui2022fast} to obtain joint confidence band for  $\beta_1^0(p)$. Similar strategy can also be employed for obtaining an asymptotic joint confidence band for the additive effect $\beta_0(p)+h(q_x(p))$, for a fixed value of $Q_X(p)=q_x(p)$. Based on the joint confidence band, it is possible to directly test for the global distributional effects $\beta(p)$ (or $h(x)$).

\newpage
\subsection{Algorithm 1 for Joint Confidence Band}
\begin{algorithm}[ht]
\caption{Joint confidence band of $\beta_1^0(p)$}
\begin{enumerate}
\label{algo 12}
\item Fit the unconstrained model and obtain the unconstrained estimator  $ \hat{\bm\psi}_{ur}=\underset{\bm\psi \in R^{K_n} }{\text{argmin}}\hspace{2 mm}  \sum_{i=1}^{n}||{\*Q_{iY}}-\^T_i\bm\psi ||_2^{2} $.
\item Fit the constrained model and obtain the constrained estimator  $  \hat{\bm\psi}_{r}=\underset{\bm\psi \in \bm\Theta_R }{\text{argmin}}\hspace{2 mm}  \sum_{i=1}^{n}||{ \*Q_{iY}}-\^T_i\bm\psi ||_2^{2}$. Obtain the constrained estimator of $\beta_1^0(p)$  as $\hat{\beta}_{1r}(p)=\rho_{K_n}(p)^{'}\hat{\bm\beta}_{1r}.$
\item  Let $\hat{\bm\Delta}_n$ be an estimate of the asymptotic covariance matrix of the unconstrained estimator given by $\hat{\bm\Delta}_n=\hat{\bm\Delta}/n= \hat{cov}(\hat{\bm\psi}_{ur})$
\item For $b = 1$ to $B$ 
\item [-]  generate $\*Z_b\sim N_{K_n}(\hat{\bm\psi}_{ur},\hat{\bm\Delta}_n)$.
\item [-] compute the projection of $\*Z_b$ as $\hat{\bm\psi}_{r,b}=\underset{\bm\psi \in \bm\Theta_R }{\text{argmin}}\hspace{2 mm} ||\bm\psi-\*Z_b||^{2}_{\hat{\*\Omega}}.$
\item [-] End For
\item For each generated sample $\hat{\bm\psi}_{r,b}$ calculate estimate of
$\beta_1^0(p)$  as $\hat{\beta}_{1r,b}(p)=\rho_{K_n}(p)^{'}\hat{\bm\beta}_{1r,b}$ ($b=1,\ldots,B$). Compute $Var(\hat{\beta}_{1r}(p))$ based on these samples.

\item For $b = 1$ to $B$ 
\item [-] calculate $u_b=\underset{p \in \mathcal{P} }{\text{max}} \frac{| \hat{\beta}_{1r,b}(p)-\hat{\beta}_{1r}(p)|}{\sqrt{Var(\hat{\beta}_{1r}(p))}}$.
\item [-] End For
\item Calculate $q_{1-\alpha}$ the $(1-\alpha)$ empirical quantile of $\{u_b\}_{b=1}^{B}$. 
\item  $100(1-\alpha)\%$ joint confidence band for $\beta_1^0(p)$ is given by 
$\hat{\beta}_{1r}(p)\pm q_{1-\alpha}\sqrt{Var(\hat{\beta}_{1r}(p))}$.
\end{enumerate}
\end{algorithm}

\section{Appendix D: Estimation of Asymptotic Covariance Matrix}
The DORQF model (8) in the paper was reformulated as 
\begin{eqnarray*}
    \*Q_{iY}&=\^T_{i} \bm{\psi} +\bm\epsilon_i, \label{dodsr:exp3},
     \end{eqnarray*}   
where $\^T_i=[\^B_0 \hspace{2mm}\^W_{i1} \hspace{2mm}\^W_{i2},\ldots,\^W_{iq} \hspace{2mm} \^S_i]$. Under suitable regularity conditions \citep{huang2004polynomial}, $\sqrt{n}(\hat{\bm\psi}_{ur}-\bm\psi^0)$ can be shown to be asymptotically distributed as $N(0,\bm\Delta)$ (also holds true for finite sample sizes if $\epsilon(p)$ is Gaussian). In reality, $\bm\Delta$ is unknown and we want to estimate $\bm\Delta$ by an estimator $\hat{\bm\Delta}$.
We derive a sandwich covariance estimator $\hat{\bm\Delta}$ corresponding to the above model. Based on the ordinary least square optimization criterion for model (11) (of the paper), the unrestricted estimator is given by $\hat{\bm\psi}_{ur}= (\^T^T\^T)^-1\^T^T\*Q_{Y}$, where $\*Q_{Y}^{T}=( \*Q_{1Y}, \*Q_{2Y},\ldots, \*Q_{nY})^T$ and $\^T=[\^T_1^{T},\^T_2^{T},\ldots,\^T_n^{T}]^T$. Hence, $Var(\hat{\bm\psi}_{ur})=(\^T^T\^T)^{-1}\^T^T\#\Sigma\^T(\^T^T\^T)^{-1}$. Here $\#\Sigma=Var(\epsilon)$, which is typically unknown. We apply an FPCA based estimation approach \citep{ghosal2022score} to estimate $\#\Sigma$.

Let us assume \citep{huang2004polynomial} the error process $\epsilon(p)$ can be decomposed as $\epsilon(p)=V(p) + w_p$, where $V(p)$ is a smooth mean zero stochastic process with covariance kernel $G(p_1,p_2)$ and $w_p$ is a white noise with variance $\sigma^2$. The covariance function of the error process is then given by $\Sigma(p_1,p_2)=cov\{\epsilon(p_1),\epsilon(p_2)\}=G(p_1,p_2) + \sigma^2 I(p_1=p_2)$. For data observed on dense and regular grid $\mathcal{P}$, the covariance matrix of the residual vector $\bm\epsilon_i$ is $\#\Sigma_{m\times m}$, the covariance kernel $\Sigma(p_1,p_2)$ evaluated on the grid $\mathcal{P}= \{p_{1}, p_{2},\ldots, p_{m} \}$. We can estimate $\Sigma(\cdot,\cdot)$ nonparametrically using functional principal component analysis (FPCA) if the original residuals $\epsilon_{ij}$ were available. Given $\epsilon_i(p_j)$s, FPCA \citep{yao2005functional} can be used to get $\hat{\phi}_k(\cdot)$, $\hat{\lambda}_k$s and $\hat{\sigma}^2$ to form an estimator of $\Sigma(p_1,p_2)$ as
$$\hat{\Sigma}(p_1,p_2)=\sum_{k=1}^{K}\hat{\lambda}_k\hat{\phi}_k(p_1)\hat{\phi}_k(p_2) + \hat{\sigma}^2 I(p_1=p_2),$$ 
where $K$ is large enough such that percent of variance explained (PVE) by the selected eigencomponents exceeds some pre-specified value such as $99\%$.

In practice, we don't have  the original residuals $\epsilon_{ij}$.
Hence we fit the unconstrained DORQF model (8) and 
and obtain the residuals $e_{ij}=Q_{iY}(p_j)-\hat{Q_{iY}}(p_j)$. Then treating $e_{ij}$ as our original residuals, we can obtain $\hat{\Sigma}(p_1,p_2)$ and $\hat{\#\Sigma}_{m\times m}$ using the FPCA approach outlined above. Then $\hat{Var}(\epsilon)=\hat{\#\Sigma}=diag\{\hat{\#\Sigma}_{m\times m},\hat{\#\Sigma}_{m\times m},\ldots,\hat{\#\Sigma}_{m\times m}\}$.  \cite{ghosal2022score} show the consistency of $\hat{\#\Sigma}$ under standard regularity conditions. Hence an consistent estimator of the covariance matrix is given by $\hat{Var}(\hat{\bm\psi}_{ur})=(\^T^T\^T)^{-1}\^T^T\hat{\#\Sigma}\^T(\^T^T\^T)^{-1}$. In particular, $\hat{\bm\Delta}_n=\hat{\bm\Delta}/n= \hat{cov}(\hat{\bm\psi}_{ur})=(\^T^T\^T)^{-1}\^T^T\hat{\#\Sigma}\^T(\^T^T\^T)^{-1}$.

\section{Appendix E: Bootstrap Test for Global Distributional Effects}
A practical question of interest in the DORQF model is to directly test for the global distributional effect of the scalar covariates $Z_j$ or test for the distributional effect of the distributional predictor $Q_X(p)$. In this section, we illustrate an nonparametric bootstrap test based on our proposed estimation method which also easily lends itself to the required shape constraints of the regression problem. In particular, we obtain the residual sum of squares of the null and  the full model and come up with the F-type test statistic defined as \begin{equation}T_D=\frac{RSS_N-RSS_F}{RSS_F}
.\label{teststat}
\end{equation}
Here $RSS_N,RSS_F$ are the residual sum of squares under the null and the full model respectively. For example, let us consider the case of testing $$H_{0} : \beta_r(p) =0 \hbox{ for all $p\in [0,1]$ \;\; versus \;\; } H_{1} : \beta_r(p) \neq 0 \hbox{  for some $p\in [0,1]$} .$$ 
Let $r=q$ without loss of generality.
The residual sum of of squares for the full model is given by $RSS_F=\sum_{i=1}^{n}||\*Q_{iY}- \^B_0\hat{\bm{\beta_0}} -\sum_{j=1}^{q}\^W_{ij} \hat{\bm{\beta_j}} -
\^S_i\hat{\bm{\theta}}||_2^{2},$ where the estimates are obtained from the optimization criterion (9) in the paper, with the constraint $\^D_F\bm\psi\geq \bm 0$ (denoting the constraint matrix for the full model as $\^D_F$). Similarly, we have $RSS_N=\sum_{i=1}^{n}||\*Q_{iY}- \^B_0\hat{\bm{\beta_0}} -\sum_{j=1}^{q-1}\^W_{ij} \hat{\bm{\beta_j}} -
\^S_i\hat{\bm{\theta}}||_2^{2},$ where the estimates are again obtained from (9) with the constraint $\^D_N\bm\psi\geq \bm 0$. Note that, in this case the constraint matrix is denoted by $D_N$ and this is essentially a submatrix of $\^D_F$ as the conditions for monotinicity in (1)-(3) (Theorem 1) for the reduced model is a subset of the original constrains for the full model. The null distribution of the test statistic $T_D$ is nonstandard, hence we use residual bootstrap to approximate the null distribution. The complete bootstrap procedure for testing the distributional effect of a scalar predictor is presented in algorithm  (\ref{algo 2}) below. Similar strategy could be employed for testing the distributional effect of a distributional predictor or multiple scalar predictors.


\begin{algorithm}[H]
\caption{Bootstrap algorithm for testing the distributional effect of a scalar predictor}
\begin{enumerate}
   \label{algo 2}
\item  Fit the full DORQF model in the paper using the optimization criterion 
\begin{equation*}
     \hat{\bm\psi}_F=\underset{\bm\psi}{\text{argmin}}\hspace{2 mm}  \sum_{i=1}^{n}||\*Q_{iY}- \^B_0\bm{\beta_0} -\sum_{j=1}^{q}\^W_{ij} \bm{\beta_j} -
    \^S_i\bm{\theta}||_2^{2} \hspace{4mm} \textit{s.t \hspace{ 4 mm} $\^D_F\bm\psi\geq \bm 0$}.
\end{equation*}
and calculate the residuals $e_i(p_l)=Q_{iY}(p_l)-\hat{Q}_{iY}(p_l)$, for $i=1,2,\ldots,n$ and $l=1,2,\ldots,m$.
\item  Fit the reduced model corresponding to $H_{0}$ (the null) and estimate the parameters using the minimization criteria, 
 \begin{equation*}
     \hat{\bm\psi}_N=\underset{\bm\psi}{\text{argmin}}\hspace{2 mm}  \sum_{i=1}^{n}||\*Q_{iY}- \^B_0\bm{\beta_0} -\sum_{j=1}^{q-1}\^W_{ij} \bm{\beta_j} -
    \^S_i\bm{\theta}||_2^{2} \hspace{4mm} \textit{s.t \hspace{ 4 mm} $\^D_N\bm\psi\geq \bm 0$}.
\end{equation*}
Denote the estimates of the distributional effects as $\hat{\beta}^{N}_{j}(p)$ for $j=0,1,\ldots,q-1$ and $\hat{h}^{N}(x)$.
\item  Compute test statistic $T_D$ (\ref{teststat}) based on these null and full model fits, denote this as $T_{obs}$.
\item  Resample B sets of bootstrap residuals $\{e^*_{b,i}(p)\}_{i=1}^{n}$ from residuals $\{e_{i}(p)\}_{i=1}^{n}$ obtained in step 1.
\item  for $b = 1$ to $B$ 
\item  Generate distributional response under the reduced DORQF model as
$$ Q^*_{b,iY}(p)=\hat{\beta}_0^{N}(p) +\sum_{j=1}^{q-1}z_{ij}\hat{\beta}^N_j(p) +\hat{h}^{N}(Q_{iX}(p))+e^*_{b,i}(p).$$
\item  Given the bootstrap data set $\{Q_{iX}(p),Q^*_{b,iY}(p), z_1,z_2,\ldots,z_q\}_{i=1}^{n}$ fit the null and the full model to compute the test statistic $T^*_b$.
\item  end for
\item  Calculate the p-value of the test as $\hat{p}=\frac{\sum_{b=1}^{B} I(T^*_b \geq T_{obs})}{B}$.
\end{enumerate}
\end{algorithm}


\section {Appendix F: Additional Simulation Scenarios}
\subsection*{Scenario A2: DORQF, Testing the effect of scalar predictor}
We consider the data generating scheme as in scenario A1 of the paper and test for the distributional effect of the scalar predictor $z_1$ using the proposed joint-confidence band based test in section 2. To this end we let $\beta_1(p)=d \times sin(\frac{\pi}{2}p)$, where the parameter $d$ controls the departure from the null hypothesis  $H_{0} : \beta_1(p) =0 \hbox{ for all $p\in [0,1]$ \;\; versus \;\; } H_{1} : \beta_1(p) \neq 0 \hbox{  for some $p\in [0,1]$}.$
The number of subject-specific measurements $L$ is set to $200$ and sample sizes $n \in \{200,300,400\}$ are considered. 
\subsection*{Scenario B: DORQF, Only distributional predictor}
We consider the following distribution on distribution regression model 
\begin{equation}
    Q_{iY}(p)=h(Q_{iX}(p))+\epsilon_i(p). \label{dodsr:sim2}
\end{equation}
The residual error process $\epsilon(p)$ is 
again generated as $\epsilon_i(p)=A_i\beta_0(p)$, $A_i\sim U(-0.5,0.5)$, which is of bounded variation. The distributional outcome $Q_{iY}(p)$ and the distributional predictor $Q_{iX}(p)$ are generated similarly as in Scenario A1. In this case, we empirically validate that the above specifications for $\epsilon_i(p)$ results in a non-decreasing quantile function $Q_{iY}(p)$.
The number of subject-specific measurements $L$ is set to $200$ and sample sizes $n \in \{200,300,400\}$ are considered. 
This scenario is used to compare the performance of the proposed DORQF method with that of the isotonic regression approach illustrated in \cite{ghodrati2022distribution}.  We consider 100 Monte-Carlo (M.C) replications from simulation scenarios B to assess the performance of the proposed estimation method. For scenario A2, 500 replicated datasets are used to assess type I error and power of the proposed testing method.

\subsection{Additional Simulation Results}
\subsection*{Performance under scenario A2:}
We assess the performance of the proposed joint confidence band based testing method in terms of estimated type I error and power calculated from the Monte-Carlo replications. We set the order of the Bernstein polynomial basis $N=3$ based on our results from previous section. The estimated power curve from the confidence band based test is displayed as a function of the parameter $d$ in Supplementary Figure S1, using a nominal level of $\alpha=0.05$. At $d=0$, the null hypothesis holds and the power corresponds to the type I error of the test. The nominal level $\alpha=0.05$ lies within its two standard error limit for all the sample sizes, illustrating that the test maintains proper size. For $d>0$, we see the power quickly increase to $1$ (higher for larger sample sizes), showing that the proposed test is able to capture departures from the null hypothesis successfully.
The estimated power curve from the nonparametric bootstrap test (and compared to the confidence band based test) is displayed in Supplementary Figure S2. It can be noticed that the confidence band based test yields a slightly higher power.

\subsection*{Performance under scenario B:}
We again consider estimation of the estimable additive effect we consider estimation of the estimable additive effect $\gamma(p)=\beta_0(p)+h(q_x(p))$ at $q_x(p)= \frac{1}{n}\sum_{i=1}^{n}Q_{iX}(p)$, which can be estimated by both the proposed DORQF (4) method and the isotonic regression method \citep{ghodrati2022distribution}. Note that true $\beta_0(p)=0$, but we include a distributional intercept in our DORQF model, nonetheless, as this information is not available to practitioners. For the isotonic regression method we directly fit the model (\ref{dodsr:sim2}) without any intercept. The performance of the estimates are compared in terms of squared Bias, variance and MSE and are reported in Table S4. We observe very similar performance of the proposed  DORQF method with the PAVA based isotnic regression method (based on optimal transport).

The estimated M.C mean for the distributional effect  $\gamma(p)$ along with their respective $95\%$ point-wise confidence intervals are displayed in Supplementary Figure S3, for the case $n=400$. Again, both the method are observed to perform a good job in capturing $\gamma(p)$. The proposed DORQF method enables conditional estimation of $\gamma(p)=\beta_0(p)+h(q_x(p))$ on the entire domain $p\in [0,1]$, where as for the isotonic regression method, interpolation is required from grid level estimates. The PAVA based isotonic regression method failed to converge in $5\%$ of the cases for sample size $n=200$, where as, this issue was not faced by our proposed method. In terms of model flexibility, the isotonic regression method do not directly accommodate scalar predictors, or a distributional intercept, and keeping these points in mind our proposed method certainly provide a uniform and flexible approach for modelling distributional outcome, in the presence of both distributional and scalar predictors.

\section{Supplementary Tables}
\begin{table}[H]
\centering
\caption{Integrated squared bias, variance and mean square error of  the estimated additive effect $\gamma(p)=\beta_0(p)+h(q_x(p))$ at $q_x(p)= \frac{1}{n}\sum_{i=1}^{n}Q_{iX}(p)$ over 100 Monte-Carlo replications, Scenario A1. }
\label{tab:my-table2}
\begin{tabular}{lllllll}
\hline
Sample Size               & \multicolumn{3}{l|}{L=200}                                         & \multicolumn{3}{l}{L=400}                                         \\ \hline
$\beta_0(p)+h(q_x(p))$ & \multicolumn{1}{l}{Bias$^2$} & \multicolumn{1}{l}{Var}    & MSE   & \multicolumn{1}{l}{Bias$^2$} & \multicolumn{1}{l}{Var}    & MSE   \\ \hline
n= 200                    & \multicolumn{1}{l}{$0.001$}  & \multicolumn{1}{l}{0.034} & 0.035 & \multicolumn{1}{l}{$6 \times 10^{-4}$}  & \multicolumn{1}{l}{0.037} & 0.037 \\ \hline
n= 300                    & \multicolumn{1}{l}{$1 \times 10^{-4}$}  & \multicolumn{1}{l}{0.031} & 0.031 & \multicolumn{1}{l}{$8.2 \times 10^{-5}$}  & \multicolumn{1}{l}{0.032} & 0.032 \\ \hline
n= 400                    & \multicolumn{1}{l}{$3 \times 10^{-4}$}  & \multicolumn{1}{l}{0.030} & 0.030 & \multicolumn{1}{l}{$2 \times 10^{-4}$}  & \multicolumn{1}{l}{0.029} & 0.029 \\ \hline
\end{tabular}

\end{table}

\begin{table}[H]
\centering
\caption{Average Wasserstein distance (standard error) between true and predicted quantile functions in the test set over 100 Monte-Carlo replications, Scenario A1.}
\vspace{3 mm}
\label{tab:my-table3}
\begin{tabular}{lll}
\hline
Sample Size (train) & L=200           & L=400           \\ \hline
n= 200      & 0.9607 (0.052) & 0.9400 (0.052) \\ \hline
n= 300      & 0.9637 (0.059) & 0.9408 (0.058) \\ \hline
n= 400      & 0.9570 (0.050) & 0.9377 (0.052) \\ \hline
\end{tabular}
\end{table}



\begin{table}[H]
\centering
\caption{Coverage of the projection-based $95\%$ joint confidence interval for $\beta_1(p)$, for various choices of the order of the Bernstein polynomial (BP) basis, scenario A1, based on $100$ M.C replications with $L=200$. Average width of the joint confidence interval is given in the parenthesis. The average choices of $N$ from cross-validation for this scenario are highlighted in bold.}
\vspace{3 mm}
\label{tab:my-tables1}
\begin{tabular}{cccc}
\hline
BP order (N) & Sample size (n=200) & Sample size (n=300) & Sample size (n=400)             \\ \hline
2              & 0.98 (1.11)     & 0.94 (0.90)      & 0.93 (0.78) \\ \hline
3            & \textbf{0.97 (1.12)}     & \textbf{0.95 (0.91)}      & \textbf{0.93 (0.79)}             \\ \hline
4            & 0.98  (1.12)   & 0.92 (0.91)       & 0.92 (0.79)              \\ \hline
\end{tabular}

\end{table}

\begin{table}[H]
\centering
\caption{Integrated squared bias, variance and mean square error of  the estimated additive effect $\gamma(p)=\beta_0(p)+h(q_x(p))$ at $q_x(p)= \frac{1}{n}\sum_{i=1}^{n}Q_{iX}(p)$ over 100 Monte-Carlo replications, Scenario B, from the DORQF method and the isotonic regression method  with PAVA \citep{ghodrati2022distribution}. }
\label{tab:my-table3}
\begin{tabular}{lllllll}
\hline
Sample Size               & \multicolumn{3}{l|}{DORQF}                                         & \multicolumn{3}{l}{PAVA}                                         \\ \hline
$\beta_0(p)+h(q_x(p))$ & \multicolumn{1}{l}{Bias$^2$} & \multicolumn{1}{l}{Var}    & MSE   & \multicolumn{1}{l}{Bias$^2$} & \multicolumn{1}{l}{Var}    & MSE   \\ \hline
n= 200                    & \multicolumn{1}{l}{0.0001}  & \multicolumn{1}{l}{0.024} & 0.024 & \multicolumn{1}{l}{$1.3 \times 10^{-5}$}  & \multicolumn{1}{l}{0.024} & 0.024 \\ \hline
n= 300                    & \multicolumn{1}{l}{0.0001}  & \multicolumn{1}{l}{0.016} & 0.016 & \multicolumn{1}{l}{$4.4 \times 10^{-6}$}  & \multicolumn{1}{l}{0.016} & 0.016 \\ \hline
n= 400                    & \multicolumn{1}{l}{0.0001}  & \multicolumn{1}{l}{0.012} & 0.012 & \multicolumn{1}{l}{$3.8 \times 10^{-5}$}  & \multicolumn{1}{l}{0.012} & 0.012 \\ \hline
\end{tabular}

\end{table}

\begin{table}[H]
\centering
\caption{Descriptive statistics of age and BMI for the complete, male and female samples in the BLSA analysis.}
\vspace{3 mm}
\label{tab:my-tabler}
\centering
\small
\begin{tabular}{cccccccc}
\hline
Characteristic     & \multicolumn{2}{c|}{Complete (n=781)} & \multicolumn{2}{c|}{Male (n=384)} & \multicolumn{2}{c|}{Female (n=397)} & P value         \\ \hline
                   & Mean          & SD  & Mean     & SD      & Mean       & SD        &                 \\ \hline
Age                & 70.17               & 9.88           & 71.37         & 9.83    & 69.01           & 9.79      & 0.0008            \\ \hline
BMI (kg/$m^2$) & 27.48               & 4.85           & 27.59         & 4.17   & 27.38           & 5.43      & $0.54$          \\ \hline
\end{tabular}

\end{table}

\begin{table}[H] \centering 
  \caption{Results from  multiple linear regression model of mean heart rate on age, sex (Male), BMI and mean activity count. Reported are the estimated fixed effects along with their standard error and P-values.}
  \label{tab1} 
\begin{tabular}{@{\extracolsep{5pt}}lccc} 
\\[-1.8ex]\hline 
\hline \\[-1.8ex] 
 & \multicolumn{3}{c}{\textit{Dependent variable : Mean heart rate}} \\ 
\cline{2-4} 
\\[-1.8ex] & Value & Std.Error & P-value \\ 
\hline \\[-1.8ex] 
Intercept & \hspace{3 mm} 89.94 & 4.543 & $<2\times10^{-16^{***}}$\\
 age & $-$0.24 & 0.038 & $8.2\times10^{-10^{***}}$ \\ 
 sex & $-3.75$ & 0.71 & $1.6\times10^{-7^{***}}$\\ 
 BMI & \hspace{3mm}0.09 & 0.075 & $0.25$\\ 
 Mean activity \hspace{5mm} & 1.98 & 0.767 & $0.0101^{***}$ \\ 
\hline \\[-1.8ex] 
Observations & 781 &  &  \\ 
Adjusted $R^2$ & 0.142 &  &  \\
\hline 
\hline \\[-1.8ex] 
\textit{Note:}  & \multicolumn{3}{r}{$^{*}$p$<$0.05; $^{**}$p$<$0.01; $^{***}$p$<$0.001} \\ 

\end{tabular} 
\end{table} 

\begin{table}[H]
\centering
\caption{Average width of $95\%$ point-wise confidence intervals from the DORQF method (Left) and the unconstrained functional regression model (Right) for the distributional effects of the predictors in the BLSA application}
\vspace{3 mm}
\label{tab:my-table3}
\begin{tabular}{lll}
\hline
Coefficient (Variable) & DORQF          & Unconstrained model          \\ \hline
$\beta_{age}(p\cdot)$ (age)     & 7.38  & 8.03  \\ \hline
$\beta_{BMI}(\cdot)$ (BMI)     & 12.51  & 13.15 \\ \hline
$\beta_{gender}(\cdot)$ (gender)    & 3.05  & 3.51  \\ \hline
$h(\cdot)$ (PA)     & 5.04  & 7.99 \\ \hline
\end{tabular}
\end{table}

\section{Supplementary Figures}
\begin{figure}[H]
\begin{center}
\includegraphics[width=.65\linewidth , height=.6\linewidth]{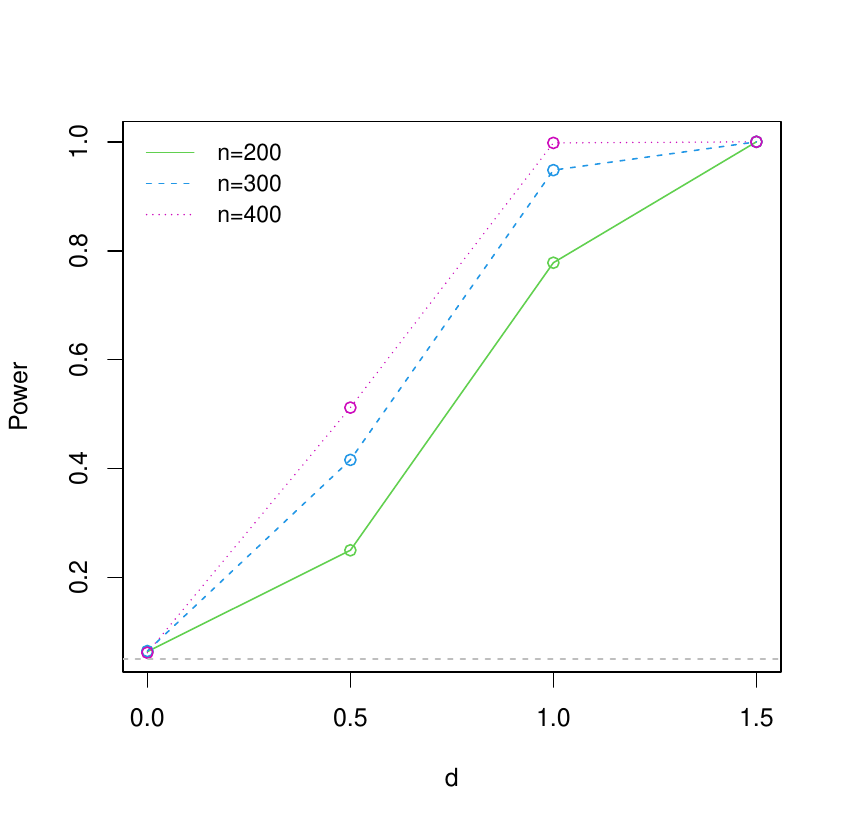} 
\end{center}
\caption{Displayed are the estimated power curves for simulation scenario A2 from the joint confidence band based test. The parameter $d$ controls the departure from the null and the power curves for $n \in\{200,300,400\}$ are shown by solid, dashed and dotted lines. The dashed horizontal line at the bottom corresponds to the nominal level of $\alpha=0.05$.}
\label{fig:fig3}
\end{figure}


\begin{figure}[H]
\begin{center}
\includegraphics[width=.8\linewidth , height=.8\linewidth]{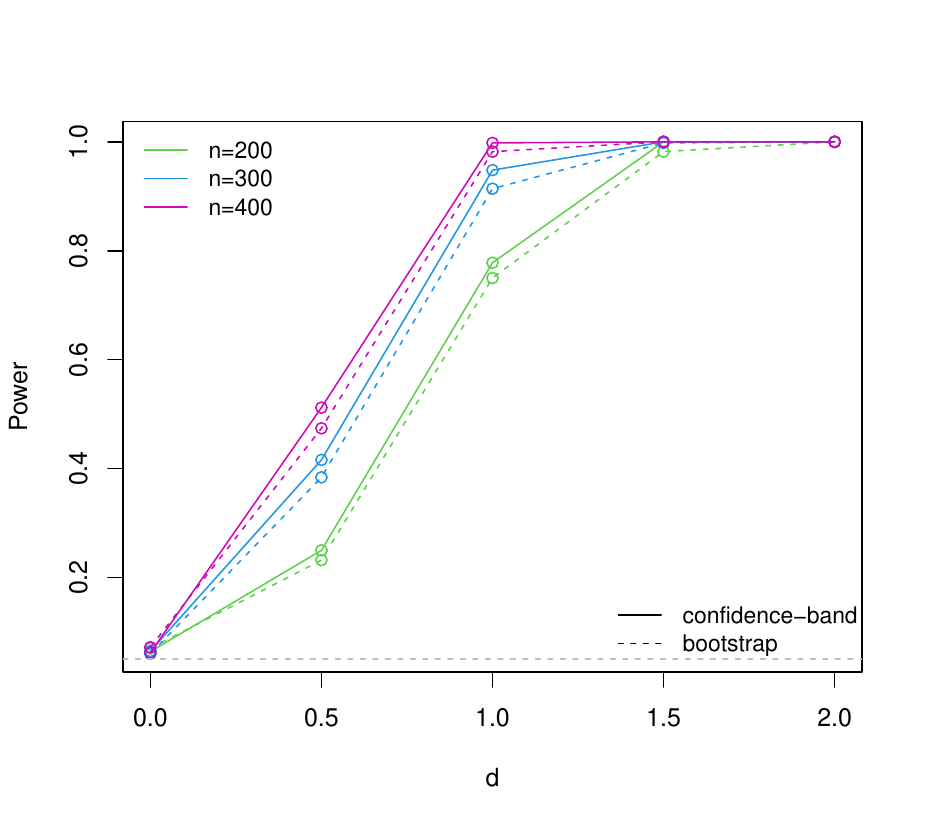} 
\end{center}
\caption{Displayed are the estimated power curves for simulation scenario A2 from the joint confidence band based (solid) and nonparametric bootstrap (dashed) test. The parameter $d$ controls the departure from the null and the power curves for $n \in\{200,300,400\}$ are shown. The dashed horizontal line at the bottom corresponds to the nominal level of $\alpha=0.05$.}
\label{fig:fig3}
\end{figure}

\begin{figure}[H]
\begin{center}
\begin{tabular}{ll}
\includegraphics[width=.5\linewidth , height=.5\linewidth]{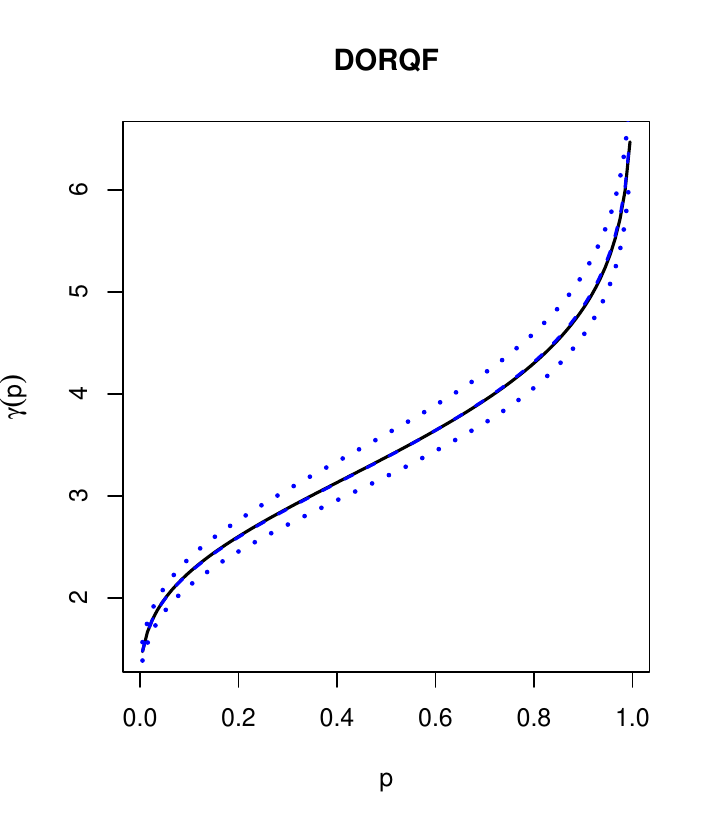} &
\includegraphics[width=.5\linewidth , height=.5\linewidth]{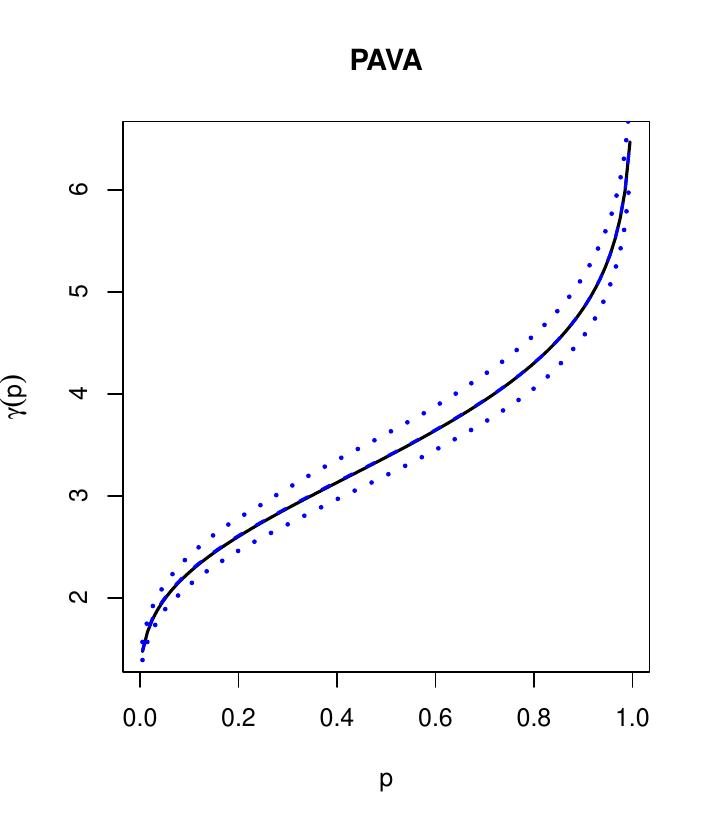}
\end{tabular}
\end{center}
\caption{Displayed are estimates of additive effect $\gamma(p)=\beta_0(p)+h(q_x(p))$ (solid) at at $q_x(p)= \frac{1}{n}\sum_{i=1}^{n}Q_{iX}(p)$ and its estimate $\hat{\gamma}(p)$ averaged over 100 M.C replications (dashed) along with point-wise $95\%$ confidence interval (dotted) for scenario B, $n=400$. Left: Estimates from the proposed DORQF method. Right: Isotonic regression method with PAVA.}
\label{fig:fig4}
\end{figure}

\begin{figure}[H]
\centering
\includegraphics[width=.9\linewidth , height=.7\linewidth]{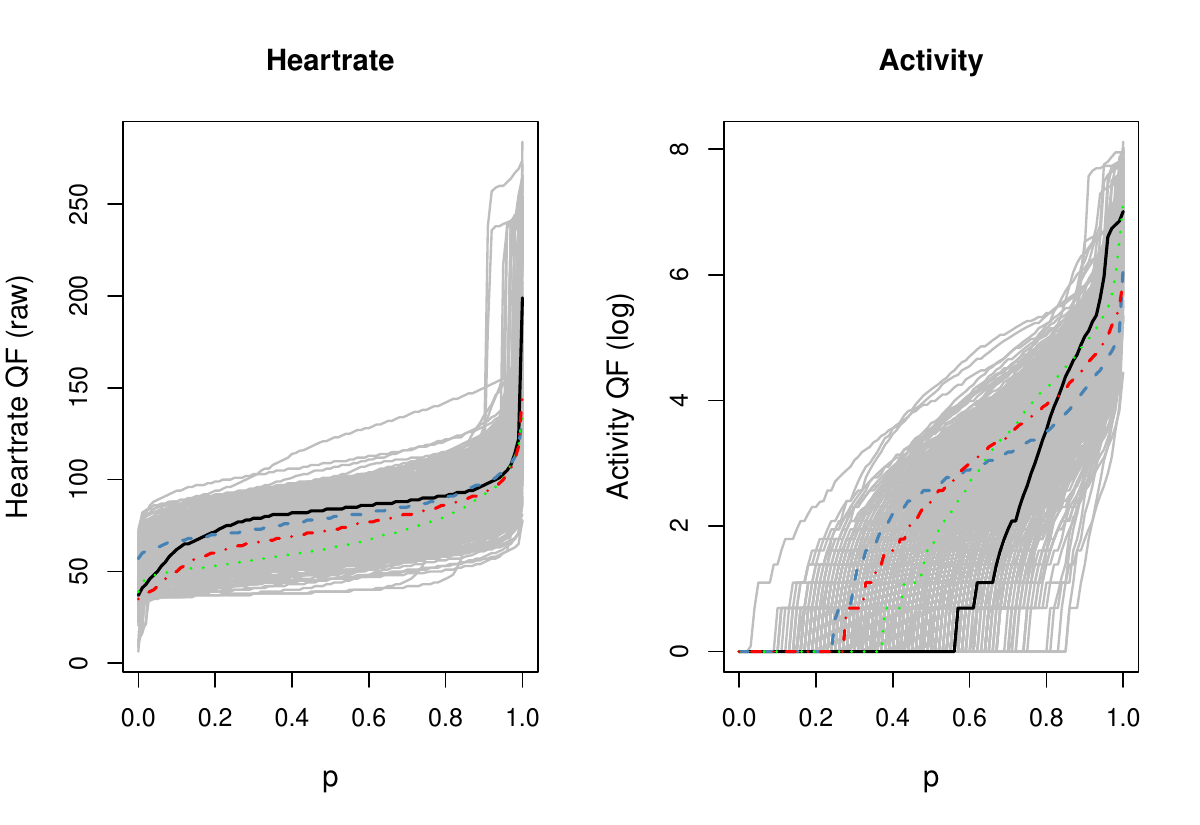}
\caption{Subject-specific quantile functions of heart rate and log-transformed activity counts during 8 a.m.- 8 p.m. period. Color profiles show four randomly chosen participants.}
\label{fig:fig1rpop}
\end{figure}

\begin{figure}[H]
\centering
\includegraphics[width=1\linewidth , height=0.8\linewidth]{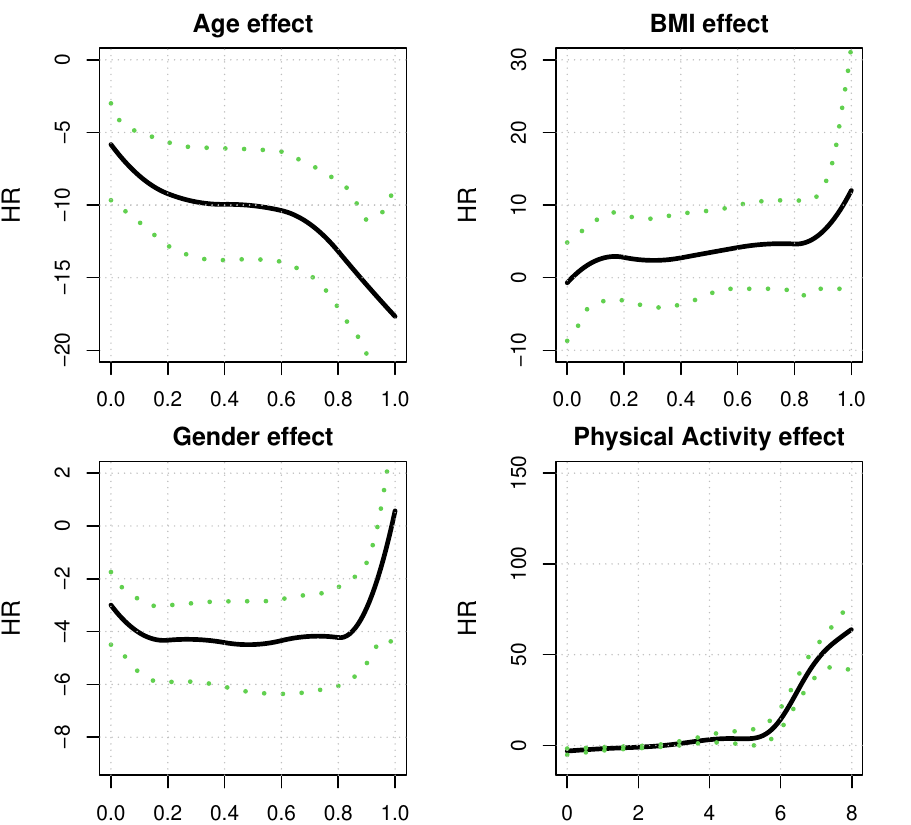}
\caption{Estimated unconstrained distributional effects (solid black) along with their bootstrapped point-wise 95$\%$ confidence intervals (dotted green) for age ($\beta_{age}(p)$), BMI ($\beta_{BMI}(p)$) and gender (Male = 1, $\beta_{gender}(p)$) on heart rate along with the estimated link function $h(\cdot)$ (bottom middle panel) between the distributions of heart rate and physical activity. The unconstrained estimates are penalized spline estimates obtained using the \texttt{"pffr"} function within the \texttt{refund} package. }
\label{fig:fig5}
\end{figure}

\newpage
\begin{figure}[H]
\centering
\includegraphics[width=1\linewidth , height=0.8\linewidth]{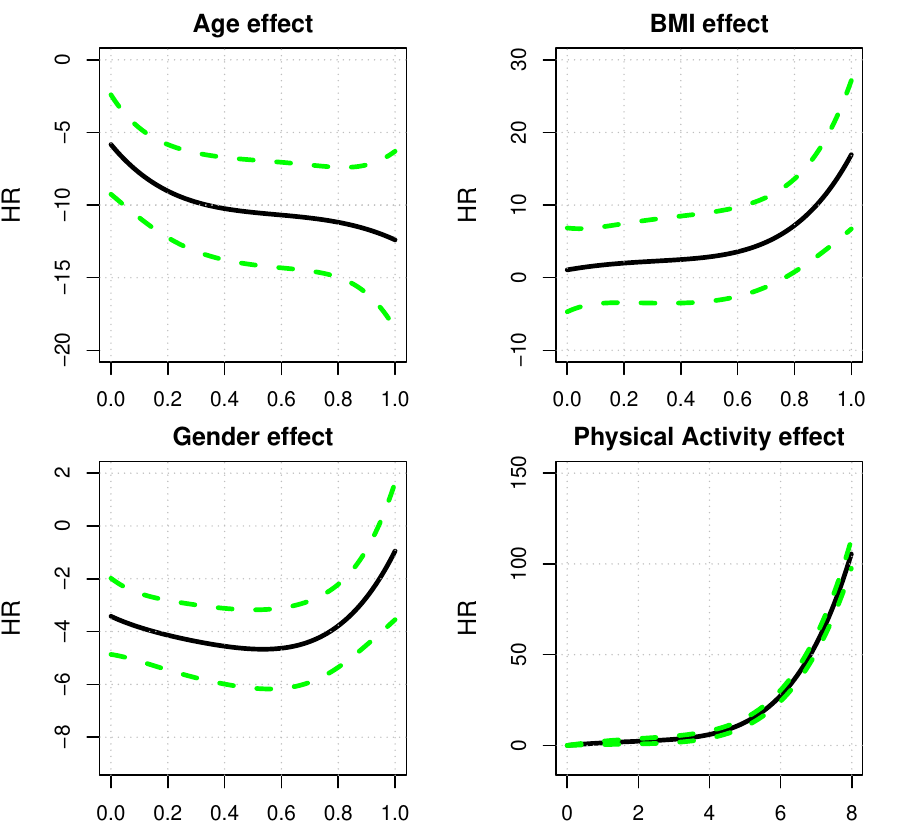}
\caption{Estimated distributional effects (solid black) along with their point-wise 95$\%$ confidence intervals (dashed green) from the proposed  DORQF model for age ($\beta_{age}(p)$), BMI ($\beta_{BMI}(p)$) and gender (Male = 1, $\beta_{gender}(p)$) on heart rate along with the estimated link function $h(\cdot)$ (bottom middle panel) between the distributions of heart rate and physical activity.}
\label{fig:fig5}
\end{figure}

\newpage

\begin{figure}[H]
\begin{center}
\includegraphics[width=.85\linewidth , height=.8\linewidth]{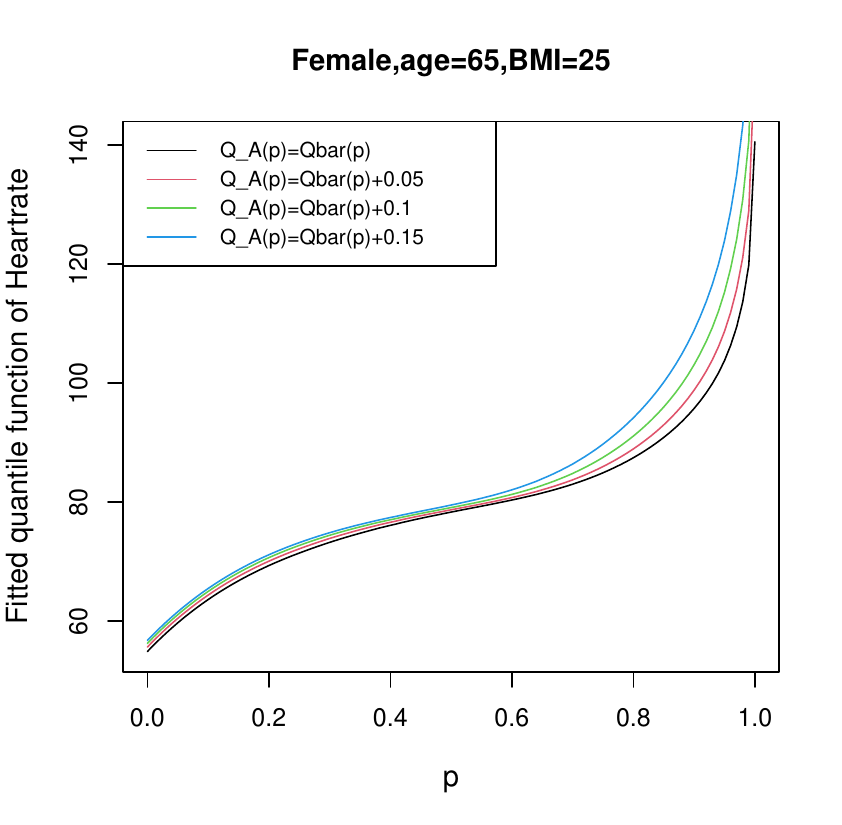} 
\end{center}
\caption{Predicted heart rate quantile functions for a female, age = 65, BMI=25 at different levels of PA quantile function.}
\label{fig:figs4}
\end{figure}

\begin{figure}[H]
\begin{center}
\includegraphics[width=1\linewidth , height=.7\linewidth]{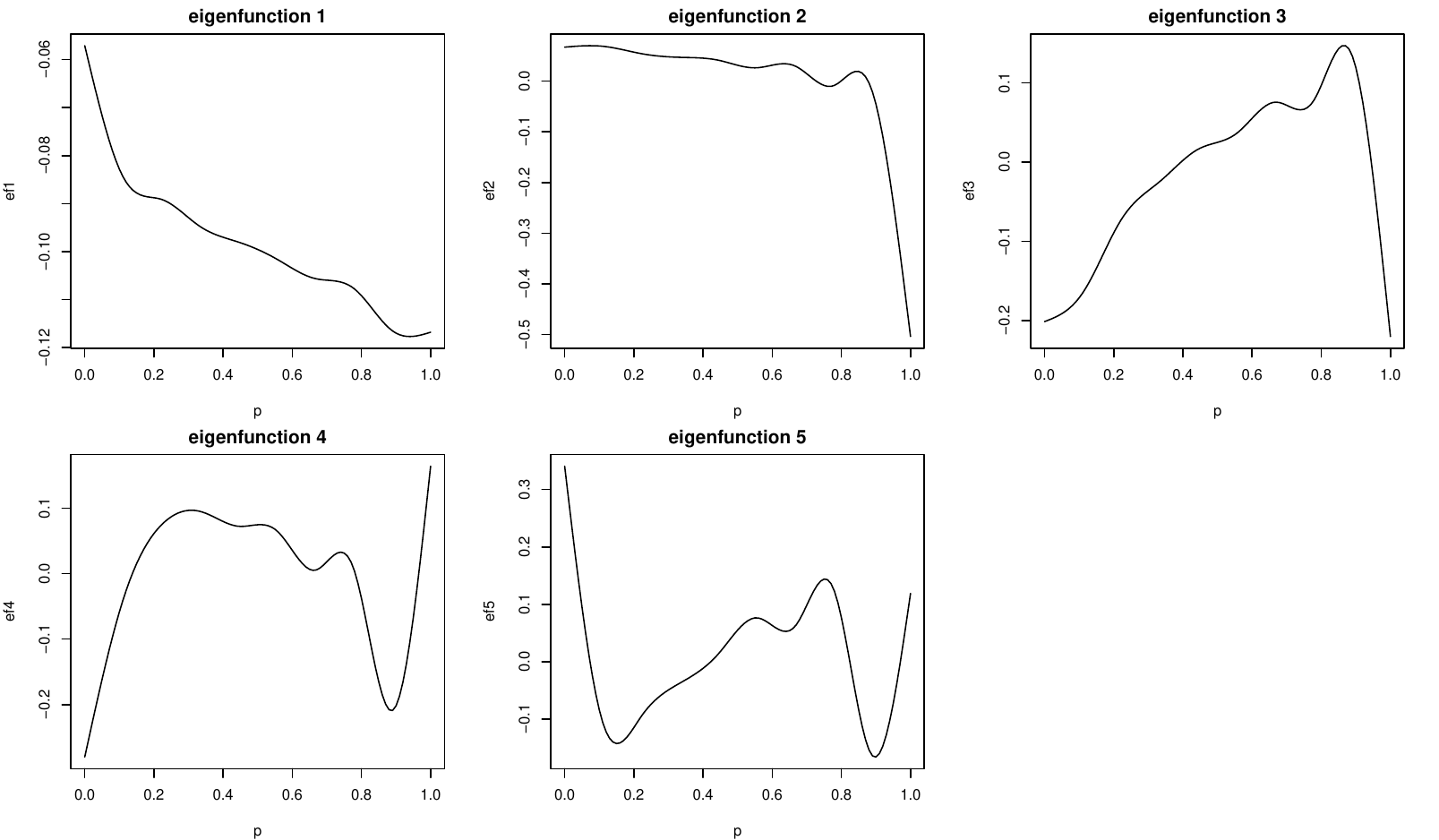} 
\end{center}
\caption{Eigenfunctions of the estimated covariance surface from the fitted residuals in the BLSA application}
\label{fig:figs6new}
\end{figure}

\begin{figure}[H]
\begin{center}
\includegraphics[width=1\linewidth , height=.7\linewidth]{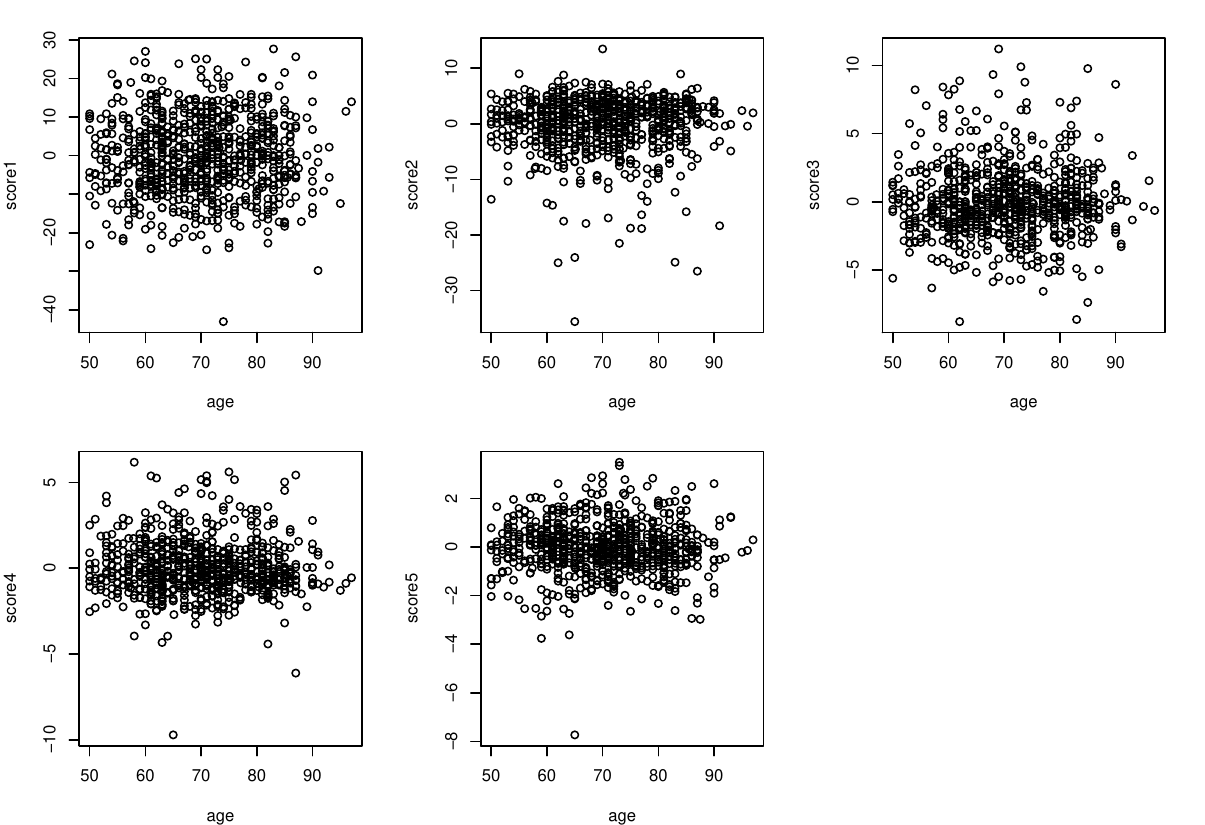} 
\end{center}
\caption{Functional principal component (FPC) scores of the fitted residuals against age in the BLSA application}
\label{fig:figs6new1}
\end{figure}

\begin{figure}[H]
\begin{center}
\includegraphics[width=1\linewidth , height=.7\linewidth]{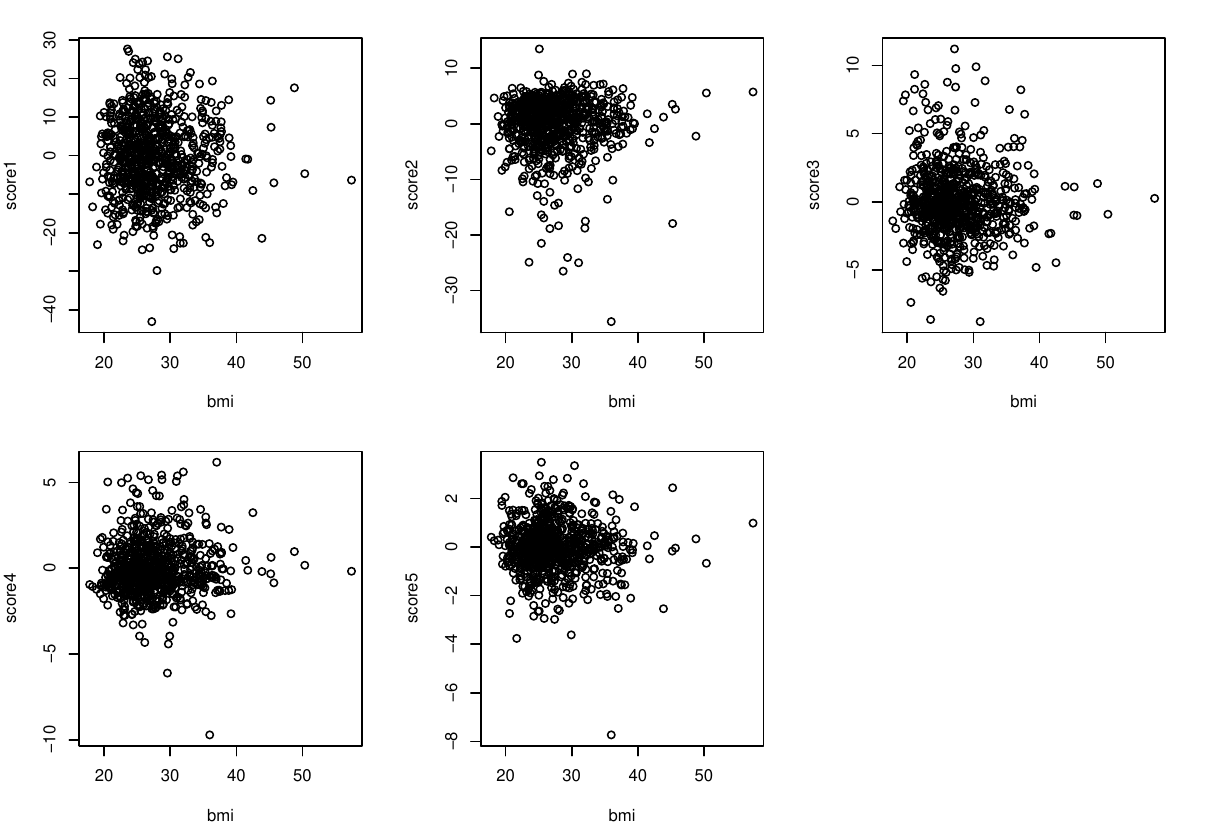} 
\end{center}
\caption{Functional principal component (FPC) scores of the fitted residuals against BMI in the BLSA application}
\label{fig:figs6new2}
\end{figure}

\begin{figure}[H]
\begin{center}
\includegraphics[width=1\linewidth , height=.7\linewidth]{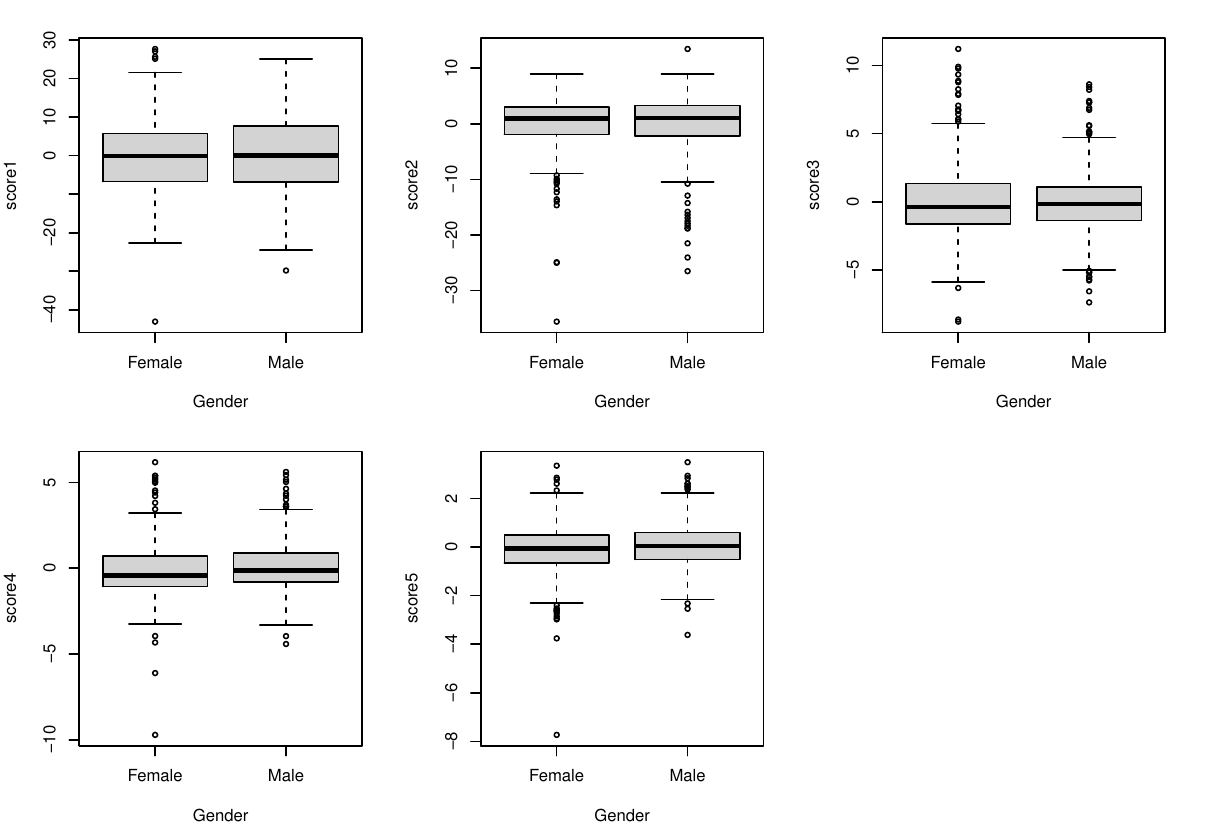} 
\end{center}
\caption{Distribution of functional principal component (FPC) scores of the fitted residuals by gender in the BLSA application}
\label{fig:figs6new3}
\end{figure}

\begin{figure}[H]
\begin{center}
\includegraphics[width=1\linewidth , height=1\linewidth]{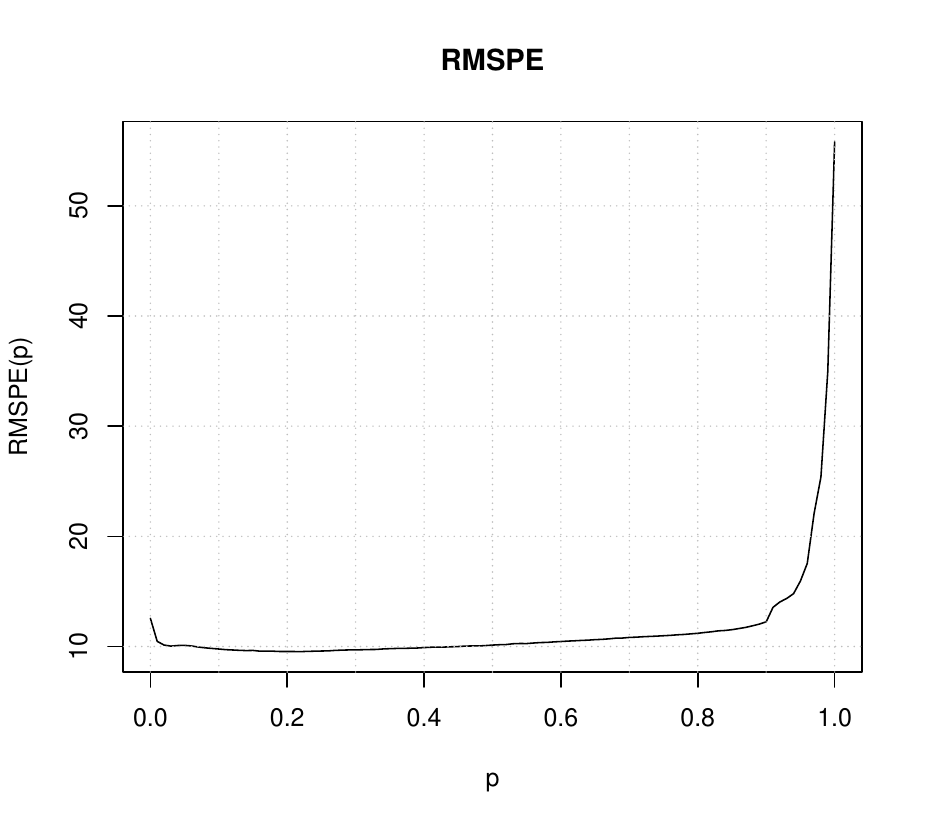} 
\end{center}
\caption{Root-mean-square prediction error (RMSPE) between the observed and predicted quantile functions in the BLSA application.}
\label{fig:figs6new4}
\end{figure}

\begin{figure}[H]
\begin{center}
\begin{tabular}{l}
\includegraphics[width=.9\linewidth , height=.44\linewidth]{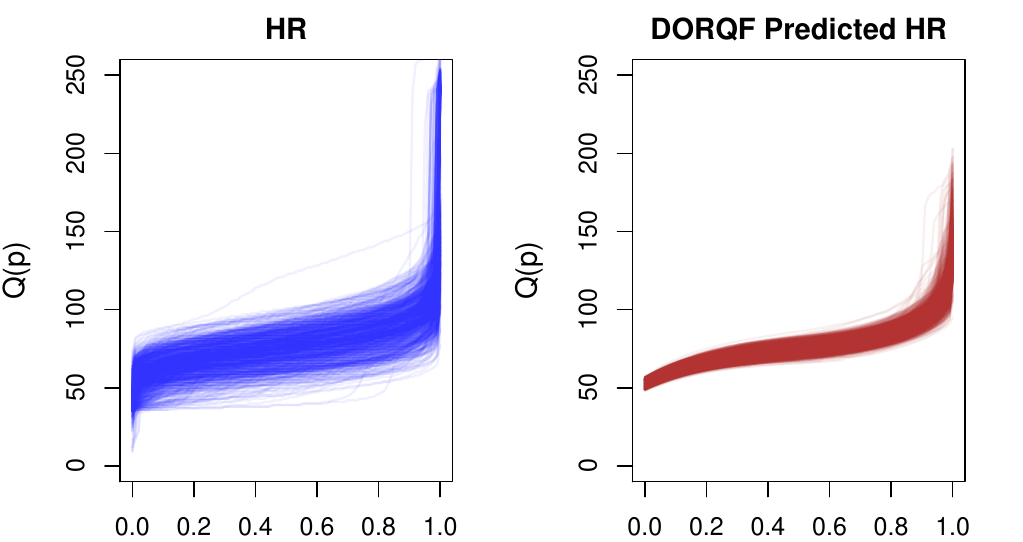} \\
\includegraphics[width=.9\linewidth , height=.5\linewidth]{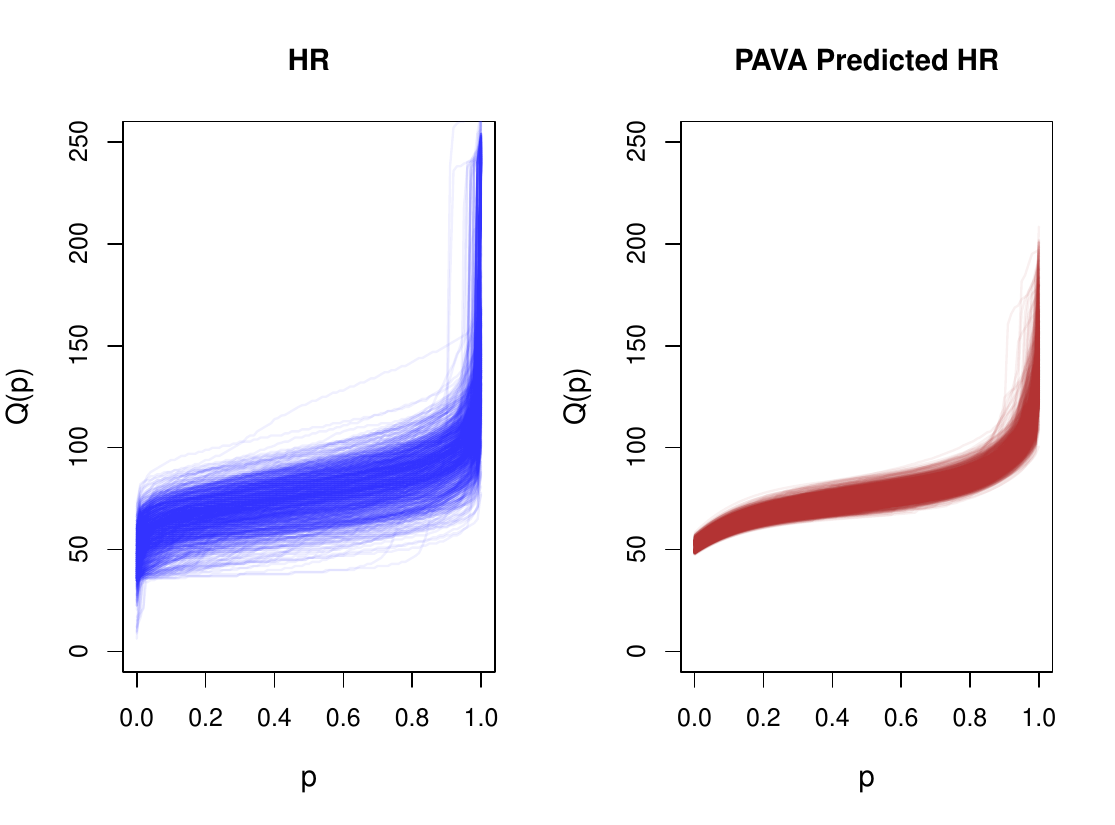}
\end{tabular}
\end{center}
\caption{Top: LOOCV predictions of quantile functions of heart rate from DORQF method based on age, sex, BMI and PA distribution. Bottom: LOOCV predictions of quantile functions of heart rate from PAVA method \citep{ghodrati2022distribution} based on PA distribution.}
\label{fig:fig6}
\end{figure}

\bibliographystyle{chicago}
\bibliography{refs}